\documentclass[journal,final]{IEEEtran}
\usepackage[T1]{fontenc}
\usepackage{verbatim}
\usepackage{amsmath}
\usepackage{amssymb}
\usepackage{graphicx}
\usepackage{tikz}
\usepackage[noadjust]{cite}
\usepackage{subcaption}

\newcommand{\partiald}[2]{\ensuremath{\frac{\partial #1}{\partial #2}}}
\newcommand\revisionD[1]{{#1}}

\makeatletter

\providecommand{\tabularnewline}{\\}

\makeatother

\usepackage{babel}
\begin{document}
\title{Sensitivity of Single-Pulse Radar Detection to Aircraft Pose Uncertainties}
\author{Mr. Austin Costley, Dr. Randall Christensen, Dr. Robert C. Leishman,
Dr. Greg Droge\thanks{This work was supported by Air Force Research Laboratory, Wright-Patterson
Air Force Base, OH.\protect \\
A. Costley is with the Electrical and Computer Engineering Department,
Utah State University, Logan, UT 84322 USA (e-mail: austin.costley@usu.edu)\protect \\
R. Christensen is with the Electrical and Computer Engineering Department,
Utah State University, Logan, UT 84322 USA (e-mail: randall.christensen@usu.edu)\protect \\
R. Leishman is with the ANT Center, Air Force Institute of Technology,
Wright-Patterson Air Force Base, OH 45433 USA (e-mail: robert.leishman@afit.edu)\protect \\
G. Droge is with the Electrical and Computer Engineering Department,
Utah State University, Logan, UT 84322 USA (e-mail: greg.droge@usu.edu)}}
\maketitle
\begin{abstract}
Mission planners for aircraft that operate in radar detection environments are often concerned the probability of detection. The probability of detection is a nonlinear function of the aircraft pose and radar position. Current path planning techniques for this application assume that the aircraft pose is deterministic. In practice, however, the aircraft pose is estimated using a navigation filter and therefore contains uncertainty. The uncertainty in the aircraft pose induces uncertainty in the probability of detection, but this phenomenon is generally not considered when path planning. This paper provides a method for combining aircraft pose uncertainty with single-pulse radar detection models to aid mission planning efforts. The method linearizes the expression for the probability of detection and three radar cross section models. The linearized models are then used to determine the variability of the probability of detection induced by uncertainty in the aircraft pose. The results of this paper validate the linearization using Monte Carlo analysis and explore the sensitivity of the probability of detection to aircraft pose uncertainty.
\end{abstract}

\section*{Nomenclature \label{sec:nomenclature}}
\addcontentsline{toc}{section}{Nomenclature}
\begin{IEEEdescription}[\IEEEusemathlabelsep\IEEEsetlabelwidth{$x_0, y_0, \psi_0, k_0$}]
\item[$P_D$] Probability of detection
\revisionD{
\item[$\bar{P}_D$] $P_D$ at nominal state
} 
\item[$P_{fa}$] Probability of false alarm
\item[$\mathcal{S}$] Signal-to-noise ratio
\item[$\sigma_r$] Radar cross section ($m^2$)
\item[$R$] Range to target ($m$)
\item[$c_r$] Culmination constant of radar parameters
\revisionD{
\item[NED] North-East-Down coordinate frame
} 
\item[$\boldsymbol{x_a}$] Aircraft state vector
\item[$\boldsymbol{\bar{x}_a}$] Nominal aircraft state vector
\item[$\delta \boldsymbol{x_a}$] Aircraft perturbation state vector  
\item[$\boldsymbol{p_a^n}$] Aircraft position vector in NED frame
\item[$\boldsymbol{\Theta}$] Aircraft Euler angle vector                  
\item[$p_{an}$, $p_{ae}$, $p_{ad}$] Aircraft position elements in NED frame
\item[$\phi_a$, $\theta_a$, $\psi_a$] Aircraft Euler angles (roll, pitch, yaw) 
\item[$\boldsymbol{p_r^n}$] Radar position vector in NED frame
\item[$p_{rn}$, $p_{re}$, $p_{rd}$] Radar position elements in NED frame
\revisionD{
\item[$\boldsymbol{p_r^b}$] Radar position vector in aircraft body frame
\item[$p_{rx}$, $p_{ry}$, $p_{rz}$] Radar position elements in aircraft body frame
}
\item[$\sigma_{rc}$] Constant RCS 
\item[$c_c$] Constant RCS parameter
\item[$\sigma_{re}$] Ellipsoid RCS 
\item[$a$, $b$, $c$] Ellipsoid RCS parameters
\item[$\sigma_{rs}$] Simple spikeball RCS 
\item[$a_s$, $b_s$] Simple spikeball RCS parameters 
\item[$\lambda$] RCS azimuth angle
\item[$\phi$] RCS elevation angle
\item[$\theta_r$] Radar detection azimuth angle 
\item[$\phi_r$] Radar detection elevation angle  
\item[$C_{xx}$] Aircraft pose covariance
\item[$C_{P_D}$] Variance of $P_D$
\item[$\sigma_{pd}$] Standard deviation of $P_D$
\end{IEEEdescription}

\section{Introduction}
\revisionD{
Manned and unmanned aircraft are often tasked with operating under threat of detection from ground-based radar systems.
Mission planners must plan paths that maintain the probability of detection below mission-specified levels. Example missions in such environments include reconnaissance \cite{Ceccarelli_micro_uav}, radar counter-measure deployment \cite{larson_path_nodate,xiao-wei_path_2010}, and combat operations \cite{kabamba_optimal_2012}. A number of factors contribute to the probability of detection. The factors include the aircraft position and orientation (pose), radar system parameters, and the physical characteristics of the aircraft (e.g. radar cross section (RCS) models).

A common assumption when modeling the detection probability is that the aircraft pose along the planned path is deterministic and known \cite{mcfarland_motion_1999,bortoff_path_2000,chandler_uav_2000,pachter_optimal_2001,moore_radar_2002,jun_path_2003,larson_path_nodate,kabamba_optimal_2012,xiao-wei_path_2010}. In practice, uncertainties in the aircraft pose enter through a variety of sources.
First, disturbances arise due to physical limitations and environmental factors (e.g. wind).
Second, the inertial navigation system on the UAV estimates the pose of the aircraft using noisy and biased sensors \cite{savage_strapdown_2000, farrell_aided_2008, grewal_global_2020}.
Considering uncertainty in the aircraft pose is particularly important for regions where global position measurements are contested, degraded, or denied. This paper develops a framework for incorporating aircraft pose uncertainty in the calculation of the probability of detection, $P_D$, for a single-pulse radar model. It is shown that pose uncertainty can be a significant source of variability in the probability of detection.

Current path planning methods use a variety of methods to evaluate the detection risk along candidate paths.
These methods include the integrated inverse range models \cite{bortoff_path_2000,chandler_uav_2000,pachter_optimal_2001,larson_path_nodate,xiao-wei_path_2010}, peak/aggregate RCS \cite{moore_radar_2002}, radar range equation \cite{mcfarland_motion_1999}, and a logistic function approximation with consolidated radar parameters \cite{kabamba_optimal_2012}.
With the exception of \cite{kabamba_optimal_2012}, these methods do not attempt to quantify the probability of detection, which is necessary for the mission planning scenario introduced above.


The target detection literature has made various developments to create high-fidelity radar detection models. The single-pulse radar model is used to determine the instantaneous probability of detection given radar parameters, radar position, and the pose of the detected aircraft.
Marcum \cite{marcum_statistical_1960} expresses target detection as a probability using a nonlinear function of the radar parameters, target radar cross section (RCS), and range to target.
Swerling \cite{swerling_probability_1954} extends the work by Marcum to include fluctuating targets for multiple-pulse detection models.
Mahafza \cite{mahafza_matlab_2003} extends the work by Marcum and Swerling to include considerations for modeling modern radar systems and provide expressions for common RCS models as a function of both the aircraft pose (position and orientation) relative to the radar. The probability of detection thus depends upon various radar parameters, the inverse of the relative range, and the relative pose of the aircraft with respect to the radar.

Despite the significant dependence upon relative aircraft pose, none of the literature mentioned above considers the effects of aircraft pose uncertainty on the probability of detection. The primary contribution of this paper is a framework for combining aircraft state uncertainty with the single-pulse radar detection models presented by Mahafza. This is accomplished by linearizing the expression for the probability of detection with respect to aircraft pose. The linearized system is then used to approximate, to the first order, the variance of the probability of detection. The results validate the linearization by running a Monte Carlo analysis and illustrate the sensitivity of the probability of detection to aircraft pose uncertainty. The resulting framework may be used by mission planners to better understand the detection risk, plan detection-aware paths, and make UAV design decisions, such as determining the required sensor quality.

This paper is organized as follows. The single-pulse radar detection and RCS models are discussed in Section \ref{sec:RadarDetectionModel}. The framework for linearizing the expression for the probability of detection and incorporating aircraft pose uncertainty is presented in Section \ref{sec:LinPd}. The results for this paper are provided in Section \ref{sec:Results} where the linearization is validated and the sensitivity of the probability of detection to aircraft pose uncertainty is presented.
}

\section{Radar Detection Model \label{sec:RadarDetectionModel}}
This section describes the radar detection model used in this work. The model includes expressions for the probability of detection, $P_D$, signal-to-noise ratio, $\mathcal{S}$, and RCS, $\sigma_r$. These quantities are functions of the aircraft pose and radar position. This section uses two RCS models from \cite{mahafza_matlab_2003} and defines the simple spikeball as an additional model. 

\subsection{Probability of Detection}
\begin{figure}
    \centering
    \includegraphics[width=\columnwidth]{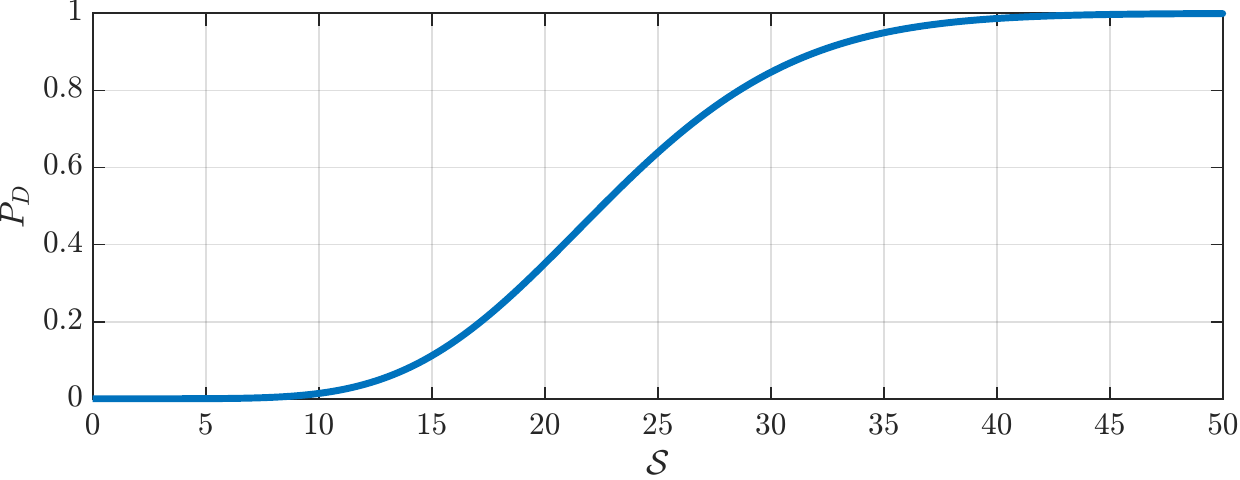}
    \caption{$P_D$ with respect to $\mathcal{S}$ for a constant $P_{fa}=1e^{-10}$. $\mathcal{S}$ is determined by the aircraft state and radar position and parameters.\label{fig:pd_snr}}
\end{figure}

\revisionD{
$P_D$ is a function of the probability of false alarm, $P_{fa}$, and the signal-to-noise ratio, $\mathcal{S}$.
$P_{fa}$ is a considered a constant for a particular radar, while $\mathcal{S}$ depends both on radar parameters and the relative target pose.
}
A common expression for $P_D$ is Marcum's Q-function \cite{mahafza_matlab_2003}. However, it contains an integral that does not have a closed form solution. 
\revisionD{
An accepted and accurate approximation to $P_D$, provided by North \cite{mahafza_matlab_2003, north_analysis_1963}, is
\begin{equation}
P_{D}\approx 0.5 \times \text{erfc}\left(\sqrt{-\ln P_{fa}}-\sqrt{\mathcal{S}+0.5}\right)\label{eq:pd_approx}
\end{equation}
where $\text{erfc}(\cdot)$
is the complementary error function given by
\begin{equation}
    \text{erfc}(z) = 1 - \frac{2}{\sqrt{\pi}} \int_0^z e^{-\zeta^2} d\zeta. \label{eq:erfc}
\end{equation}
Fig. \ref{fig:pd_snr} shows an example graph of $P_D$ with respect to $\mathcal{S}$.
} 

A general expression for the signal-to-noise ratio is given by 
\begin{eqnarray}
\mathcal{S}&=&c_r\frac{\sigma_r}{kR^4}\label{eq:SNR}
\end{eqnarray}
where $k$ is Boltzmann's constant $(1.38\times10^{-23} \; J/^{\circ}K)$ and $c_r$ is a radar constant that is a function of radar parameters such as power, aperture area, noise figure, and loss factor and is dependent on the type of radar being modeled. The specific radar parameters are not critical to the development of this paper so they will be lumped into a single radar constant.

\revisionD{
It is important to note here that both the RCS, $\sigma_{r}$,
and range, $R$, are functions of the pose of
the aircraft, which are depicted in Fig. \ref{fig:radar_xy}.
Let the aircraft position in the NED frame, $\boldsymbol{p_a^n}$, and orientation, $\boldsymbol{\Theta_a}$, be defined by
}
\begin{eqnarray}
    \boldsymbol{p_a^n} &=& \begin{bmatrix} p_{an} & p_{ae} & p_{ad} \end{bmatrix}^\intercal \label{eq:pa} \\
    \boldsymbol{\Theta_a} &=& \begin{bmatrix} \phi_a & \theta_a & \psi_a \end{bmatrix}^\intercal \label{eq:theta_a}
\end{eqnarray}
where $\boldsymbol{\Theta_a}$ is a vector of Euler angles for the roll, pitch, and yaw of the aircraft \cite{beard_randy_small_2012}.
Thus, \eqref{eq:SNR} is expressed more explicitly
as 
\begin{equation}
\mathcal{S}\left(\boldsymbol{p_{a}^n},\boldsymbol{\Theta_a}\right)=c_r\frac{\sigma_{r}\left(\boldsymbol{p_{a}^n},\boldsymbol{\Theta_a}\right)}{kR\left(\boldsymbol{p_{a}^n}\right)^{4}}.\label{eq:SNR_exp}
\end{equation}
and the range to the radar is given by
\begin{eqnarray}
R(\boldsymbol{p_{a}^n}) & = & ||\boldsymbol{p_{a}^n}-\boldsymbol{p_{r}^n}||_{2} \label{eq:range}
\end{eqnarray}
where $\boldsymbol{p_{r}^n}$ represents the position of the radar in the NED frame with
\begin{equation}
\boldsymbol{p_r^n} = \begin{bmatrix} p_{rn} & p_{re} & p_{rd}\end{bmatrix}^\intercal. 
\end{equation}
In this work the radar position is assumed to be deterministic and known. 

The RCS representation is vehicle specific and the RCS model can range from a simple constant to a complex spikeball. The RCS representations used in this work are described in the following subsection.

\subsection{Radar Cross Section Models \label{sec:rcs_models}}
\begin{figure}
    \centering
    \tikzset{every picture/.style={line width=0.75pt}} 

\begin{tikzpicture}[x=0.75pt,y=0.75pt,yscale=-1,xscale=1]

\draw    (146,241) -- (146,133) ;
\draw [shift={(146,131)}, rotate = 90] [color={rgb, 255:red, 0; green, 0; blue, 0 }  ][line width=0.75]    (10.93,-3.29) .. controls (6.95,-1.4) and (3.31,-0.3) .. (0,0) .. controls (3.31,0.3) and (6.95,1.4) .. (10.93,3.29)   ;
\draw    (146,241) -- (264,241) ;
\draw [shift={(266,241)}, rotate = 180] [color={rgb, 255:red, 0; green, 0; blue, 0 }  ][line width=0.75]    (10.93,-3.29) .. controls (6.95,-1.4) and (3.31,-0.3) .. (0,0) .. controls (3.31,0.3) and (6.95,1.4) .. (10.93,3.29)   ;
\draw  [color={rgb, 255:red, 128; green, 128; blue, 128 }  ,draw opacity=1 ][fill={rgb, 255:red, 128; green, 128; blue, 128 }  ,fill opacity=1 ] (263.17,105.53) -- (320.08,86.24) -- (323.29,95.71) -- (266.38,115.01) -- cycle ;
\draw  [color={rgb, 255:red, 128; green, 128; blue, 128 }  ,draw opacity=1 ][fill={rgb, 255:red, 128; green, 128; blue, 128 }  ,fill opacity=1 ] (316.95,92.58) .. controls (316.06,89.96) and (317.46,87.13) .. (320.08,86.24) .. controls (322.69,85.35) and (325.53,86.75) .. (326.42,89.37) .. controls (327.31,91.98) and (325.91,94.82) .. (323.29,95.71) .. controls (320.68,96.6) and (317.84,95.2) .. (316.95,92.58) -- cycle ;
\draw  [color={rgb, 255:red, 128; green, 128; blue, 128 }  ,draw opacity=1 ][fill={rgb, 255:red, 128; green, 128; blue, 128 }  ,fill opacity=1 ] (308.1,95.19) -- (291.93,132.17) -- (288.41,122.93) -- (298.68,99.45) -- (275.63,89.38) -- (271.81,79.32) -- cycle ;
\draw  [color={rgb, 255:red, 128; green, 128; blue, 128 }  ,draw opacity=1 ][fill={rgb, 255:red, 128; green, 128; blue, 128 }  ,fill opacity=1 ] (276.17,109.95) -- (268.87,123.66) -- (259.73,96.14) -- (274.13,103.8) -- cycle ;
\draw  [dash pattern={on 4.5pt off 4.5pt}]  (293.23,100.62) -- (293.23,62.62) ;
\draw [shift={(293.23,60.62)}, rotate = 90] [color={rgb, 255:red, 0; green, 0; blue, 0 }  ][line width=0.75]    (10.93,-3.29) .. controls (6.95,-1.4) and (3.31,-0.3) .. (0,0) .. controls (3.31,0.3) and (6.95,1.4) .. (10.93,3.29)   ;
\draw  [dash pattern={on 4.5pt off 4.5pt}]  (293.23,100.62) -- (331.23,100.62) ;
\draw [shift={(333.23,100.62)}, rotate = 180] [color={rgb, 255:red, 0; green, 0; blue, 0 }  ][line width=0.75]    (10.93,-3.29) .. controls (6.95,-1.4) and (3.31,-0.3) .. (0,0) .. controls (3.31,0.3) and (6.95,1.4) .. (10.93,3.29)   ;
\draw    (293.23,100.62) -- (323.14,89.93) -- (362.12,76.01) ;
\draw [shift={(364,75.33)}, rotate = 160.34] [color={rgb, 255:red, 0; green, 0; blue, 0 }  ][line width=0.75]    (10.93,-3.29) .. controls (6.95,-1.4) and (3.31,-0.3) .. (0,0) .. controls (3.31,0.3) and (6.95,1.4) .. (10.93,3.29)   ;
\draw  [draw opacity=0] (293.99,83.13) .. controls (301.4,83.46) and (307.58,88.68) .. (309.57,95.73) -- (293.23,100.62) -- cycle ; \draw   (293.99,83.13) .. controls (301.4,83.46) and (307.58,88.68) .. (309.57,95.73) ;
\draw    (293.23,100.62) -- (200,200) ;
\draw  [draw opacity=0] (199.61,180) .. controls (199.74,180) and (199.87,180) .. (200,180) .. controls (205.39,180) and (210.46,181.42) .. (214.83,183.92) -- (200,210) -- cycle ; \draw   (199.61,180) .. controls (199.74,180) and (199.87,180) .. (200,180) .. controls (205.39,180) and (210.46,181.42) .. (214.83,183.92) ;
\draw  [draw opacity=0] (342.2,83.06) .. controls (344.25,88.54) and (345.37,94.45) .. (345.37,100.62) .. controls (345.37,128.85) and (322.02,151.73) .. (293.23,151.73) .. controls (279.68,151.73) and (267.33,146.66) .. (258.06,138.34) -- (293.23,100.62) -- cycle ; \draw   (342.2,83.06) .. controls (344.25,88.54) and (345.37,94.45) .. (345.37,100.62) .. controls (345.37,128.85) and (322.02,151.73) .. (293.23,151.73) .. controls (279.68,151.73) and (267.33,146.66) .. (258.06,138.34) ;
\draw  [color={rgb, 255:red, 0; green, 0; blue, 0 }  ,draw opacity=1 ][fill={rgb, 255:red, 0; green, 0; blue, 0 }  ,fill opacity=1 ] (200,190) -- (210,200) -- (200,210) -- (190,200) -- cycle ;
\draw  [fill={rgb, 255:red, 0; green, 0; blue, 0 }  ,fill opacity=1 ] (290.73,100.62) .. controls (290.73,99.24) and (291.85,98.12) .. (293.23,98.12) .. controls (294.61,98.12) and (295.73,99.24) .. (295.73,100.62) .. controls (295.73,102) and (294.61,103.12) .. (293.23,103.12) .. controls (291.85,103.12) and (290.73,102) .. (290.73,100.62) -- cycle ;
\draw  [dash pattern={on 4.5pt off 4.5pt}]  (200,200) -- (200,142) ;
\draw [shift={(200,140)}, rotate = 90] [color={rgb, 255:red, 0; green, 0; blue, 0 }  ][line width=0.75]    (10.93,-3.29) .. controls (6.95,-1.4) and (3.31,-0.3) .. (0,0) .. controls (3.31,0.3) and (6.95,1.4) .. (10.93,3.29)   ;
\draw  [dash pattern={on 4.5pt off 4.5pt}]  (200,200) -- (248,200) ;
\draw [shift={(250,200)}, rotate = 180] [color={rgb, 255:red, 0; green, 0; blue, 0 }  ][line width=0.75]    (10.93,-3.29) .. controls (6.95,-1.4) and (3.31,-0.3) .. (0,0) .. controls (3.31,0.3) and (6.95,1.4) .. (10.93,3.29)   ;
\draw    (293.23,100.62) -- (305.31,133.46) ;
\draw [shift={(306,135.33)}, rotate = 249.8] [color={rgb, 255:red, 0; green, 0; blue, 0 }  ][line width=0.75]    (10.93,-3.29) .. controls (6.95,-1.4) and (3.31,-0.3) .. (0,0) .. controls (3.31,0.3) and (6.95,1.4) .. (10.93,3.29)   ;

\draw (301,59.4) node [anchor=north west][inner sep=0.75pt]    {$ \begin{array}{l}
\psi _{a}\\
\end{array}$};
\draw (200,160.4) node [anchor=north west][inner sep=0.75pt]    {$ \begin{array}{l}
\theta _{r}\\
\end{array}$};
\draw (336,131.4) node [anchor=north west][inner sep=0.75pt]    {$\lambda $};
\draw (127,122) node [anchor=north west][inner sep=0.75pt]   [align=left] {N};
\draw (277,242) node [anchor=north west][inner sep=0.75pt]   [align=left] {E};
\draw (181,211) node [anchor=north west][inner sep=0.75pt]   [align=left] {radar};
\draw (227,61.4) node [anchor=north west][inner sep=0.75pt]  [font=\scriptsize]  {$\begin{bmatrix}
p_{an}\\
p_{ae}\\
p_{ad}
\end{bmatrix}$};
\draw (157,161.4) node [anchor=north west][inner sep=0.75pt]  [font=\scriptsize]  {$\begin{bmatrix}
p_{rn}\\
p_{re}\\
p_{rd}
\end{bmatrix}$};
\draw (352,59.4) node [anchor=north west][inner sep=0.75pt]  [font=\scriptsize]  {$b_{x}$};
\draw (311,121.4) node [anchor=north west][inner sep=0.75pt]  [font=\scriptsize]  {$b_{y}$};

\end{tikzpicture}    
\caption{
    Graphical representation of the quantities used in the radar detection model.\label{fig:radar_xy}
    }
\end{figure}
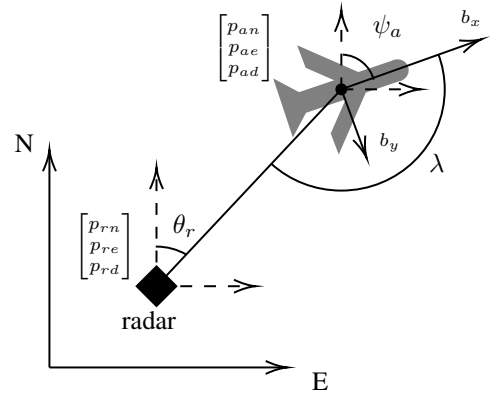

RCS models provide an expression for the RCS value as a function of the azimuth and elevation angles of the radar detection vector in the body frame of the aircraft. The body frame $x$ and $y$ axes are shown as $b_x$ and $b_y$ in Fig. \ref{fig:radar_xy}. The body frame $z$ axis points out of the bottom of the aircraft.
\revisionD{
The position of the radar in the body frame of the aircraft is given by
\begin{equation}
\boldsymbol{p_r^b} = \begin{bmatrix} p_{rx} & p_{ry} & p_{rz} \end{bmatrix}^\intercal.
\end{equation}
The vector $\boldsymbol{p_r^b}$ is calculated using the aircraft pose and radar position by
\begin{equation}
    \boldsymbol{p_r^b} = R_n^b \left(\boldsymbol{p_r^n}-\boldsymbol{p_a^n}\right)
\end{equation}
where $R_n^b$ is the direction cosine matrix formed by the ZYX Euler angle sequence \cite{beard_randy_small_2012} given by
\begin{align}
    R_n^b = &\left[\begin{matrix}  C{\psi_a} C{\theta_a} & -C{\phi_a} S{\psi_a} + C{\psi_a} S{\phi_a} S{\theta_a}  \\
                C{\theta_a} S{\psi_a} & C{\phi_a} C{\psi_a} + S{\phi_a} S{\psi_a} S{\theta_a}  \\
                -S{\theta_a} & C{\theta_a} S{\phi_a} \end{matrix}\right.\nonumber \\
        & \qquad \qquad \qquad \qquad \left.\begin{matrix}
        S{\phi_a} S{\psi_a} + C{\phi_a}C{\psi_a}S{\theta_a} \\
        -C{\psi_a} S{\phi_a} + C{\phi_a} S{\psi_a} S{\theta_a} \\
        C{\phi_a} C{\theta_a} \end{matrix}\right] \label{eq:DCM_zyx}
\end{align}
and $\textrm{S}\cdot$ and $\textrm{C}\cdot$ are the $\sin(\cdot)$ and $\cos(\cdot)$ functions.
}

The RCS azimuth angle is the angle from the body frame $x$ axis to the projection of the radar detection vector into the $x$-$y$ plane of the body frame. The RCS elevation angle $\phi$ is the angle from the $x$-$y$ plane in the body frame of the aircraft to the radar detection vector with a positive angle towards the bottom of the aircraft. The RCS azimuth and elevation angles are given by
\begin{eqnarray}
    \lambda &=& \arctan\left(\frac{p_{ry}}{p_{rx}}\right)
\end{eqnarray}
and
\begin{eqnarray}
    \phi &=& \arctan\left(\frac{p_{rz}}{\sqrt{(p_{rx})^2+(p_{ry})^2}}\right).
\end{eqnarray}

\begin{figure*}
    \bigskip
    \begin{subfigure}{0.24\textwidth}
        \includegraphics[width=\textwidth]{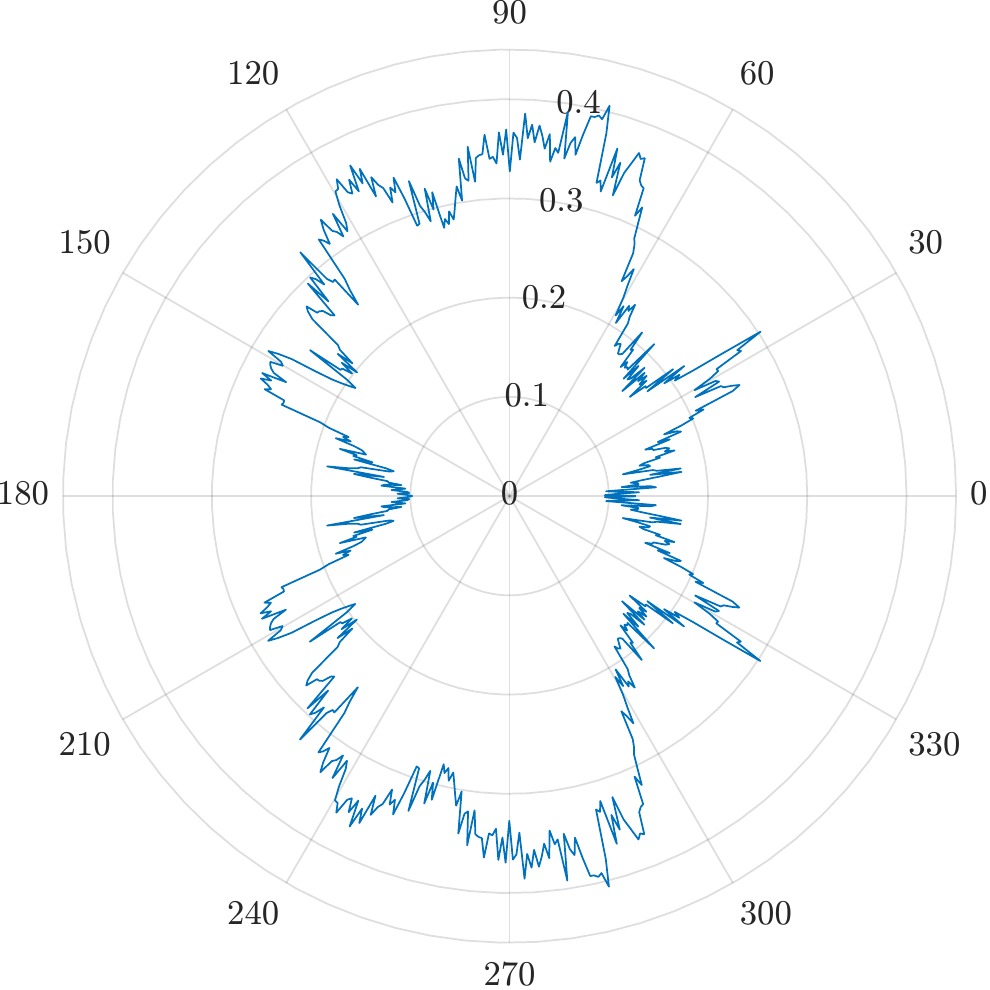}
        \caption{Complex Spikeball\label{fig:fuzz_rcs}}
    \end{subfigure}
        \begin{subfigure}{0.24\textwidth}
            \includegraphics[width=\textwidth]{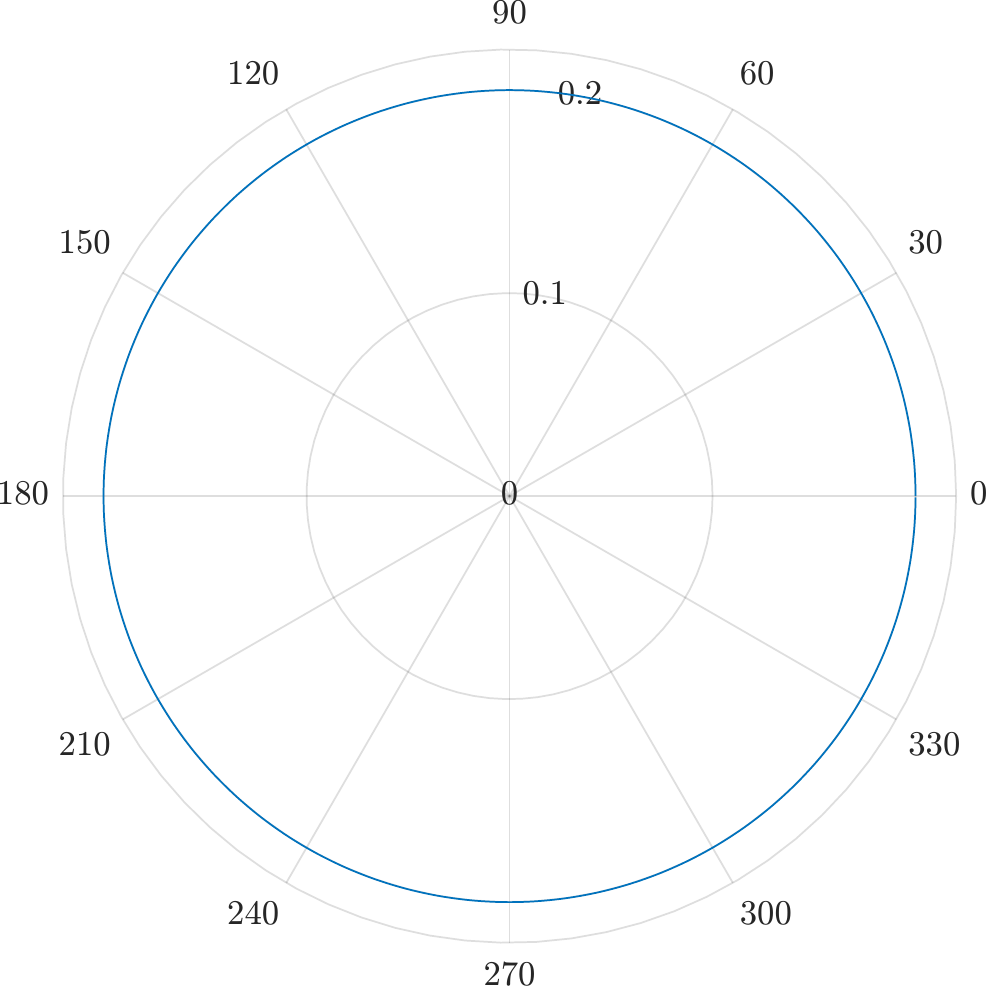}
            \caption{Constant \label{fig:constant_rcs}}
        \end{subfigure}
        \bigskip
        \begin{subfigure}{0.24\textwidth}  
            \includegraphics[width=\textwidth]{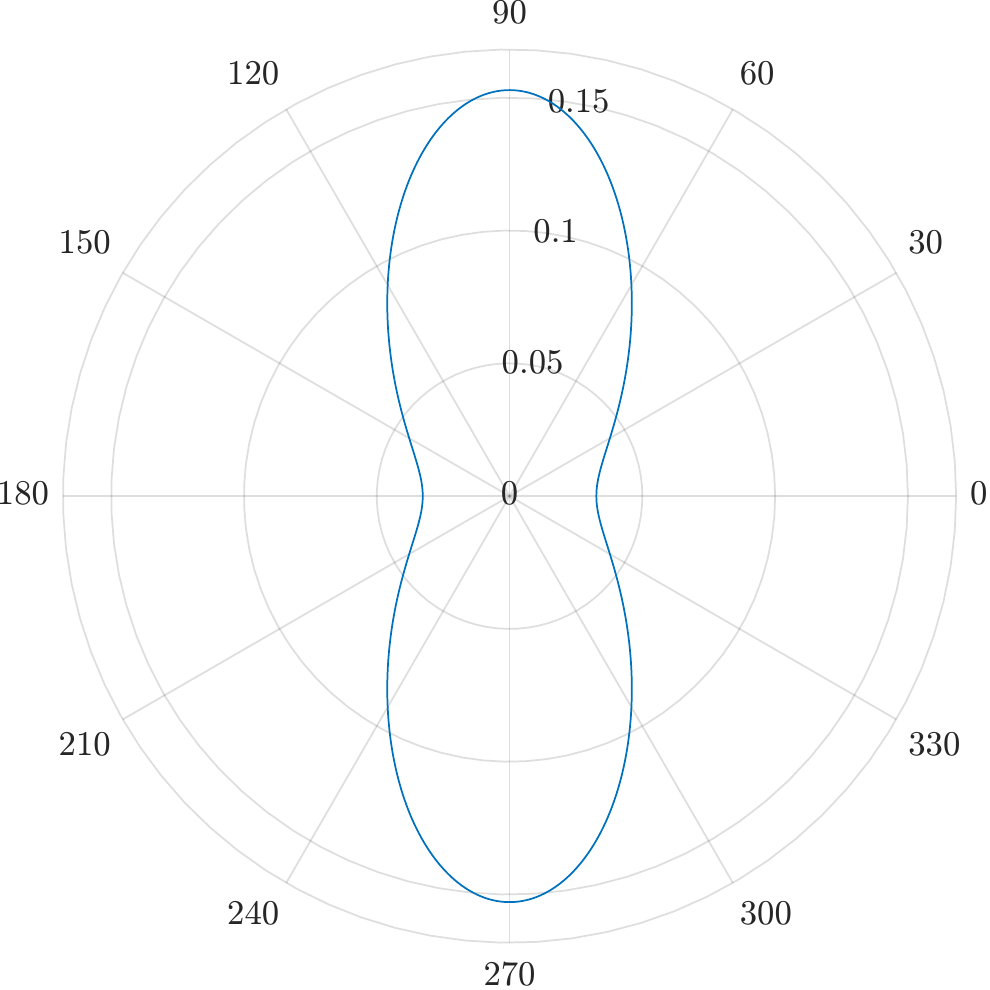}
            \caption{Ellipsoid \label{fig:ellipsoid_rcs}
            }
        \end{subfigure}
        \bigskip
        \begin{subfigure}{0.24\textwidth}    
            \includegraphics[width=\textwidth]{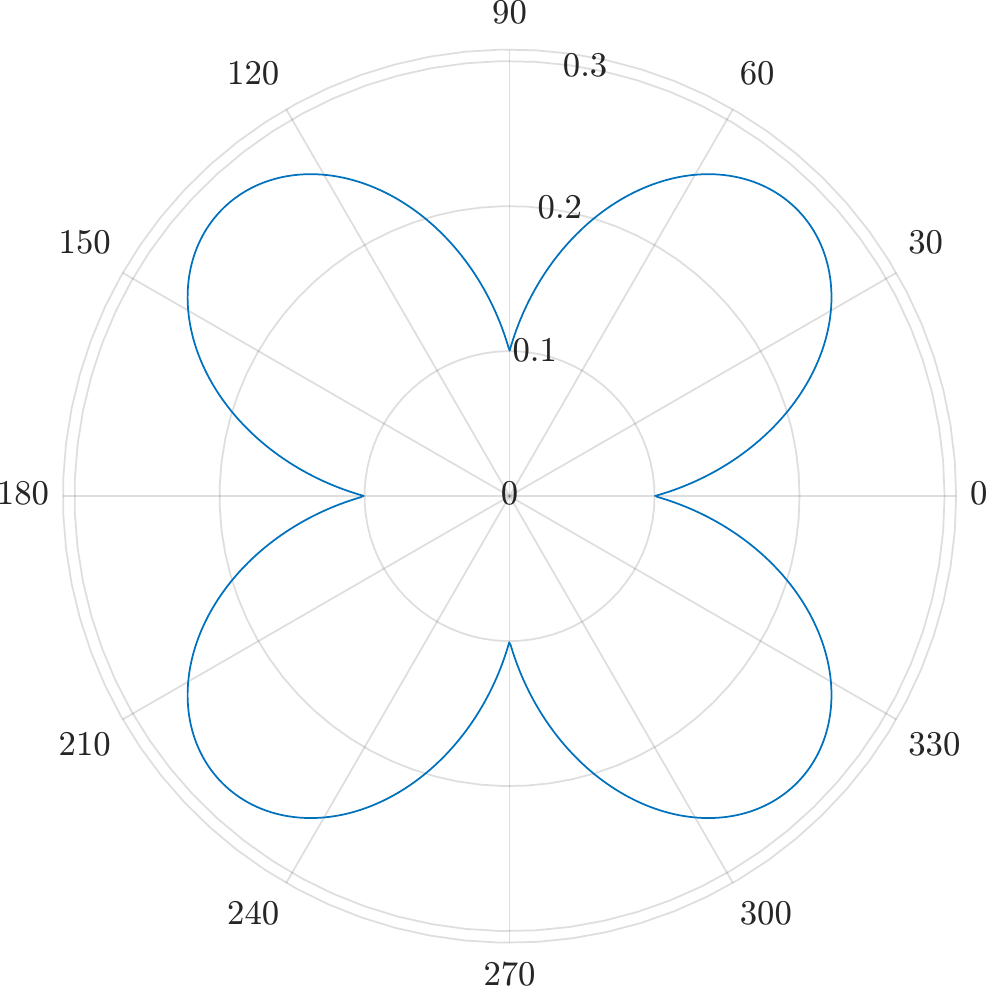}
            \caption{Simple Spikeball\label{fig:spikeball_rcs}}
        \end{subfigure}
        \vspace{-2\baselineskip}
    \caption{
            Examples of four RCS models as a function of $\lambda$ where the aircraft nose is pointed towards the $0$ degree line. (a) shows the highest fidelity model and is typically obtained through radar measurements. (b) shows the constant RCS model that is independent of aircraft pose. (c) and (d) show the ellipsoid and simple spikeball RCS models that are dependent on radar azimuth angle $\lambda$. The ellipsoid model used in this work is a 3D ellipsoid model that is also
     dependent on the radar elevation angle $\phi$ (not pictured).
     }
    \label{fig:rcs_models}%
\end{figure*}

A common graphical representation of an RCS model is a polar plot, where the angle is either the RCS azimuth or elevation angle and the radius represents the RCS value. Fig. \ref{fig:rcs_models} provides example polar plots for four radar cross section models as functions of the RCS azimuth angle. The complex spikeball (Fig. \ref{fig:fuzz_rcs}) is the highest fidelity model and is generally obtained through radar measurements and data gathering. Path planners typically use simplified models such as those in Figs. \ref{fig:constant_rcs}-\ref{fig:spikeball_rcs}. This work will use the constant, ellipsoid, and simple spikeball RCS models. The remainder of this section presents equations for computing the RCS for these models as a function of the RCS azimuth and elevation angles.

The first RCS model used in this work is the constant where
\begin{equation}
\sigma_{rc} = c_c.
\end{equation}
The constant, $c_c$, is chosen to be a conservative representation of the RCS of the aircraft. This model is the most simple and is independent of the aircraft pose. A polar plot of a constant RCS model is shown in Fig. \ref{fig:constant_rcs}.

The second RCS model is the ellipsoid. The equation for the RCS of an ellipsoid \cite{mahafza_matlab_2003,kabamba_optimal_2012} represents a 3-dimensional surface given by
\begin{align}
\sigma_{re}=\frac{\pi\left(abc\right)^{2}}{\left(\left(a \,\textrm{S}\lambda \,\textrm{C}\phi\right)^{2}+\left(b \,\textrm{S}\lambda\, \textrm{S}\phi\right)^{2}+\left(c\, \textrm{C}\lambda\right)^{2}\right)^{2}} \label{eq:rcs_ellipsoid}
\end{align}
where $a$, $b$, and $c$, are the length of the ellipsoid axes.
A polar plot of the ellipsoid RCS with respect to $\lambda$ is shown in Fig. \ref{fig:ellipsoid_rcs}. Note that the ellipsoid model is also a function of the elevation angle, $\phi$.

The third RCS model is known as 
a ``simple spikeball.'' The term comes from the shape of the plot in 
polar coordinates. An example of a simple spikeball with four
lobes is shown in Fig. \ref{fig:spikeball_rcs}.
The expression for a simple spikeball RCS is given by
\begin{eqnarray}
\sigma_{rs} & = & \left|a_s\sin\left(\frac{n}{2}\lambda\right)\right|+b_s\label{eq:spikeball_rcs}
\end{eqnarray}
where $a_s$ determines the amplitude of the lobes, $b_s$ is the minimum RCS value, and $n$ determines the number of lobes.
\revisionD{
Note that the simple spikeball RCS model does not depend on $\phi$.
}

\section{Linearized Probability of Detection \label{sec:LinPd}}
The aircraft pose is represented by a Gaussian distributed random variable with a mean $\boldsymbol{x_a} = \begin{bmatrix}\boldsymbol{p_a^n} & \boldsymbol{\Theta_a}\end{bmatrix}^\intercal$ and covariance $C_{xx}$. The prior sections have shown that $P_D$ is a nonlinear function of the aircraft state. Hence, variability in the aircraft pose induces variability in the probability of detection. The expected $P_D$ (i.e. the mean) and the uncertainty (i.e. variance) are approximated by linearizing \eqref{eq:pd_approx} about the aircraft state.

The combination of \eqref{eq:pd_approx} and \eqref{eq:SNR_exp} provides
an expression for $P_D$ as a function of the
aircraft pose, $\boldsymbol{x_a}$,  
\revisionD{
which can be expressed as a nominal pose, $\boldsymbol{\bar{x}_a}$, with a perturbation, $\delta \boldsymbol{x_a}$, as
}
\begin{equation}
    \revisionD{
    \boldsymbol{x_a} = \boldsymbol{\bar{x}_a} + \delta \boldsymbol{x_a}.
    }
\end{equation}
In general, $P_D$ is a nonlinear function with respect to the pose of the
detected aircraft.
However, variations in $P_D$ can be approximated by linearizing Eqs. \eqref{eq:pd_approx}, \eqref{eq:SNR_exp}, and \eqref{eq:range} about a nominal operating point using a Taylor series expansion to obtain
\revisionD{
\begin{align}
\delta P_{D} & \approx \frac{\partial P_D}{\partial \mathcal{S}}\left(\frac{\partial \mathcal{S}}{\partial R}\frac{\partial R}{\partial\boldsymbol{x_{a}}}+\frac{\partial \mathcal{S}} {\partial\sigma_{r}}\frac{\partial\sigma_{r}}{\partial\boldsymbol{x_{a}}}\right) \Bigg\rvert_{\boldsymbol{\bar{x}_a}} \delta\boldsymbol{x_{a}} \label{eq:pd_partial_long}\\
 & \approx A_P\delta\boldsymbol{x_{a}}. \label{eq:pd_partial_short}
\end{align}
where $\delta P_D$ is the perturbation of $P_D$ due to the perturbation of the aircraft state.
Then the variance of $P_D$ due to aircraft pose uncertainty is computed using a similarity transform as
}
\revisionD{
\begin{align}
C_{P_D} & = E\left[\delta P_{D}\delta P_{D}^{\intercal}\right]\\
 & = A_P C_{xx} A_P^{\intercal} \label{eq:pd_var}
\end{align}
and the standard deviation is given by 
\begin{equation}
    \sigma_{pd} = \sqrt{C_{P_D}} \label{eq:pd_sig}.
\end{equation}
These equations show that $A_P$ must be calculated to compute the variability of $P_D$. 
}
The following subsections will define the five partial derivatives from \eqref{eq:pd_partial_long} that are used to compute $A_P$. 

\subsection{Partial Derivative of $P_D$ with Respect to $\mathcal{S}$}
The partial derivative of $P_D$ with respect to $\mathcal{S}$ is computed by combining \eqref{eq:pd_approx} and \eqref{eq:erfc} to obtain
\begin{eqnarray}
    P_{D} & \approx & 0.5-\frac{1}{\sqrt{\pi}}\int_{0}^{W}\exp\left(-\zeta^{2}\right)d\zeta
\end{eqnarray}
where
\begin{eqnarray}
W & = & \sqrt{-\ln P_{fa}}-\sqrt{\mathcal{S}+0.5}.
\end{eqnarray}
Proceed by computing the derivative of $P_{D}$ with respect to $W$ by applying the
fundamental theorem of calculus%
\begin{align}
\frac{\partial P_{D}}{\partial W} & =\frac{-\exp\left(-W^2\right)}{\sqrt{\pi}}
\end{align}
The partial derivative of $W$ with respect to $\mathcal{S}$
is given by%
\begin{align*}
\frac{\partial W}{\partial\mathcal{S}} & =\frac{-1}{2\sqrt{\mathcal{S}+0.5}}.
\end{align*}
It follows that
\begin{align}
\frac{\partial P_{D}}{\partial\mathcal{S}} & =\frac{\partial P_{D}}{\partial W}\frac{\partial W}{\partial\mathcal{S}}\nonumber \\
 & =\frac{\exp\left(-W^2\right)}{2\sqrt{\pi}\sqrt{\mathcal{S}+0.5}}.\label{eq:dp_dsnr}
\end{align}

\subsection{Partial Derivative of $\mathcal{S}$ with Respect to Range}
The partial derivative of the signal-to-noise ratio \eqref{eq:SNR_exp} with respect to range is given by%
\begin{align}
\frac{\partial\mathcal{S}}{\partial R} & =-c_r\frac{4 \sigma_r}{kR^{5}}.
\end{align}

\subsection{Partial Derivative of Range with Respect to Aircraft State}
The range or distance between the aircraft and the radar is given by \eqref{eq:range}.
The partial derivative of range with respect to the aircraft position is given by

\begin{align}
\frac{\partial R}{\partial\boldsymbol{x_{a}}} & = \begin{bmatrix}\frac{\partial}{\partial\boldsymbol{p_{a}}}||\boldsymbol{p_{a}-p_{r}}||_{2} & \boldsymbol{0_{1x3}}\end{bmatrix}\\
 & =\begin{bmatrix}\frac{\left(\boldsymbol{p_{a}-p_{r}}\right)^\intercal}{||\boldsymbol{p_{a}-p_{r}}||_{2}} & \boldsymbol{0_{1x3}}\end{bmatrix}.
\end{align}

\subsection{Partial Derivative of $\mathcal{S}$ with Respect to Radar Cross Section}
The partial derivative of the
signal-to-noise ratio \eqref{eq:SNR_exp} with respect to the RCS is given
by
\begin{align}
\frac{\partial\mathcal{S}}{\partial\sigma_{r}} & = c_r\frac{1}{kR^{4}}.
\end{align}

\subsection{Partial Derivative of RCS with Respect to Aircraft State}
The partial derivative of the RCS with respect to the aircraft state, $\frac{\partial \sigma_r}{\partial \boldsymbol{x_a}}$, is dependent on the choice of RCS model. The lengthy derivation of the partial derivatives of the three RCS models presented in Section \ref{sec:rcs_models} are provided in Appendix \ref{ap:rcs_linearization}.

\subsection{Combining Expressions for $A_P$}
Finally, the partial derivatives calculated in the preceding subsections are combined to obtain an expression for $A_P$ as
\begin{eqnarray}
    A_P &=& \begin{bmatrix} 
            -c_r\frac{2\text{exp}\left(-W^2\right)\left(\boldsymbol{p_{a}-p_{r}}\right)^\intercal}{kR^{6}\sqrt{\pi}\sqrt{\mathcal{S}+0.5}} & \boldsymbol{0_{1x3}}
            \end{bmatrix} \\
           & & +
            c_r\frac{\exp\left(-W^2\right)}{2kR^{4}\sqrt{\pi}\sqrt{\mathcal{S}+0.5}} \frac{\partial \sigma_r}{\partial \boldsymbol{x_a}}
\end{eqnarray}
where $\frac{\partial \sigma_r}{\partial \boldsymbol{x_a}}$ depends on the chosen RCS model as defined in Appendix \ref{ap:rcs_linearization}. The resulting expression for $A_p$ is used to compute the variance of $P_D$ with respect to the aircraft state uncertainty using \eqref{eq:pd_sig}. Section \ref{sec:Results} provides results that validate the linearization presented in this section and illustrates the sensitivity of $P_D$ to aircraft state uncertainty.

\section{Results \label{sec:Results}}
The results for this paper are separated into two sections. First, results are presented to show the validity of the linearization of the radar detection and RCS models. Second, the sensitivity of $P_D$ to the aircraft state uncertainty is presented. Both sections provide results for the three RCS models presented in Section \ref{sec:rcs_models} for three levels of aircraft state uncertainty (low, medium, high). The radar model used in the results is a surveillance radar with parameters matching examples from \cite{mahafza_matlab_2003}. Table \ref{tab:RadarParamsResults} provides all of the parameter values used in this section.

\begin{table}
    \begin{centering}
    \caption{Radar parameters for used in results section \label{tab:RadarParamsResults}}
    \par\end{centering}
    \centering{}%
    {\renewcommand{\arraystretch}{1.25}
    \begin{tabular}{c|c|l}
    \normalsize{\textbf{Param}} & \normalsize{\textbf{Value}} & \normalsize{\textbf{Description}}\tabularnewline
    \hline 
    $c_c$ & 0.2 $\textrm{m}^2$ & RCS for constant model \tabularnewline
    $a$ & 0.25 $\textrm{m}$ & Ellipsoid RCS forward axis length \tabularnewline
    $b$ & 0.15 $\textrm{m}$ & Ellipsoid RCS side axis length\tabularnewline
    $c$ & 0.17 $\textrm{m}$ & Ellipsoid RCS up axis length \tabularnewline
    $a_s$ & 0.2 $\textrm{m}^2$ & Spikeball RCS lobe amplitude \tabularnewline
    $b_s$ & 0.15 $\textrm{m}^2$ & Spikeball RCS minimum \tabularnewline
    $n$ & 4 & Number of lobes \tabularnewline
    L ${\sigma_{pa}}$ & 0.1 $\textrm{m}$ & Low position state std. dev.\tabularnewline
    M ${\sigma_{pa}}$ & 10 $\textrm{m}$ & Medium position state std. dev.\tabularnewline
    H ${\sigma_{pa}}$ & 100 $\textrm{m}$ & High position state std. dev.\tabularnewline
    L ${\sigma_{ang}}$ & 0.1 deg.& Low Euler angle state std. dev.\tabularnewline
    M ${\sigma_{ang}}$ & 1 deg.& Medium Euler angle state std. dev.\tabularnewline
    H ${\sigma_{ang}}$ & 2 deg.& High Euler angle state std. dev.\tabularnewline
    $\bar{\psi}
    _a$ & 0 deg. & Aircraft course angle \tabularnewline
    $\bar{p}_{ad}$ & $-3$ km & Aircraft position along "down" axis \tabularnewline
    $\boldsymbol{p_r^n}$ & $\boldsymbol{0_{3\times1}}$ $\textrm{m}$ & Radar position vector (NED) \tabularnewline
    $c_r$ & $167.4$ J$\textrm{m}^2/^{\circ}$K  & Lumped radar constant \tabularnewline
    $P_{fa}$ & $1.7e^{-4}$ & Probability of false alarm \tabularnewline
    $R$ & $650$ km & Range to aircraft
    \end{tabular}
    }
    \end{table}

\subsection{Linearization Validation}
The linearization described in Section \ref{sec:LinPd} provides a first order approximation to variations in the nonlinear radar detection model due to variations in aircraft state. The validity of this approximation is dependent on the operating point $\boldsymbol{\bar{x}_a}$, the RCS model, and the pose uncertainty $C_{xx}$ and is assessed in this section.

\revisionD{
The linearization validation will be illustrated using Monte Carlo analysis for a scenario where the radar is at the origin of the NED frame ($\boldsymbol{p_r^n}=\boldsymbol{0_{3\times1}}$) and the aircraft position is rotated around the radar at a nominal range.
In this approach, the aircraft is considered at a series of nominal poses that are perturbed with random samples according to the state uncertainty level. 
The $k^{th}$ nominal aircraft pose is given by
\begin{equation}
    \boldsymbol{\bar{x}_a}[k] = \begin{bmatrix} R\sin(\theta_r[k]) & R\cos(\theta_r[k]) & \bar{p}_{ad} & \boldsymbol{0_{1\times3}} \end{bmatrix}^\intercal,
\end{equation}
where $\theta_r[k]$ ranges from 0--180 degrees in increments of 0.5 degrees.
The nominal range, $R = 650$ km, ensures that $P_D$ is near $0.5$ for some values of $\theta_r$ given the RCS models defined in Table \ref{tab:RadarParamsResults} and $\bar{p}_{ad}=-3$ km is within the operating altitude for tactical unmanned aircraft \cite{weibel_safety_2005}.  
The nominal state for Monte Carlo run $i$ is perturbed using
\begin{equation}
    \boldsymbol{x_{a,i}}[k] = \boldsymbol{\bar{x}_{a,i}}[k] + \boldsymbol{w_i}[k]
\end{equation}
where $\boldsymbol{w_i}[k]$ is sampled as a zero-mean Gaussian distributed random vector with a covariance matrix given by
\begin{equation}
    C_{xx} = \begin{bmatrix} \sigma_{pa}\mathbf{I_{3 \times 3}}  & \mathbf{0} \\ \mathbf{0} & \sigma_{ang}\mathbf{I_{3 \times 3}}  \end{bmatrix}.
\end{equation}
The perturbed states are used to calculate $P_D[k]$ using \eqref{eq:pd_approx} for each $\theta_r[k]$ value.
The collection of $P_D$ values over the range of $\theta_r$ make up a single Monte Carlo run.
The full Monte Carlo analysis consisted of 1000 runs.
In addition to $P_D$ calculated from the perturbed states, $\bar{P}_D$ is calculated with \eqref{eq:pd_approx} using the nominal aircraft state, $\boldsymbol{\bar{x}_a}$.
} 

\begin{figure}
    \centering{}
    \includegraphics[width=1\columnwidth]{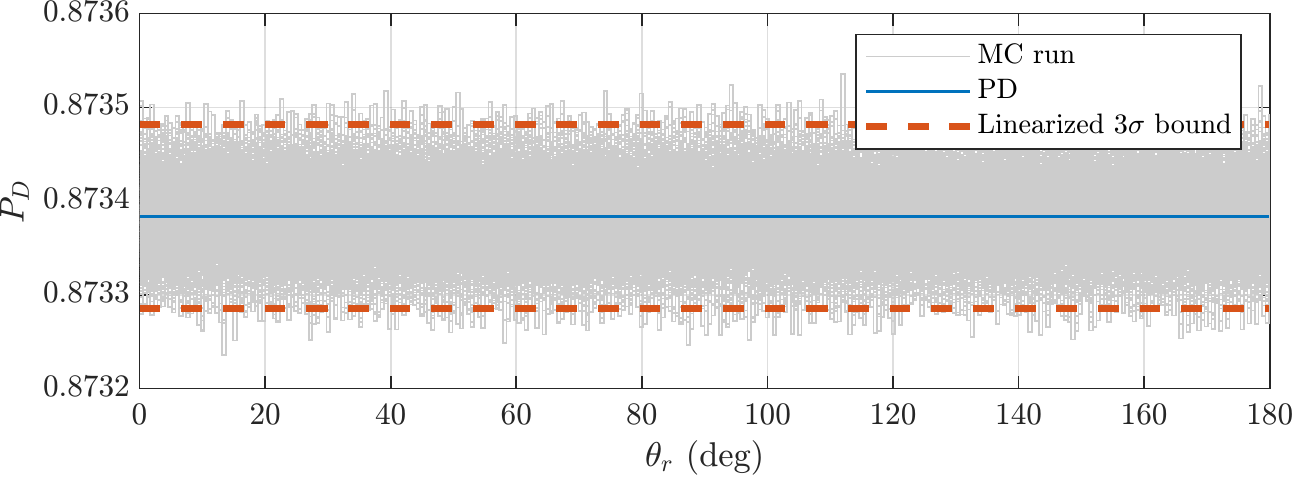}
    \includegraphics[width=1\columnwidth]{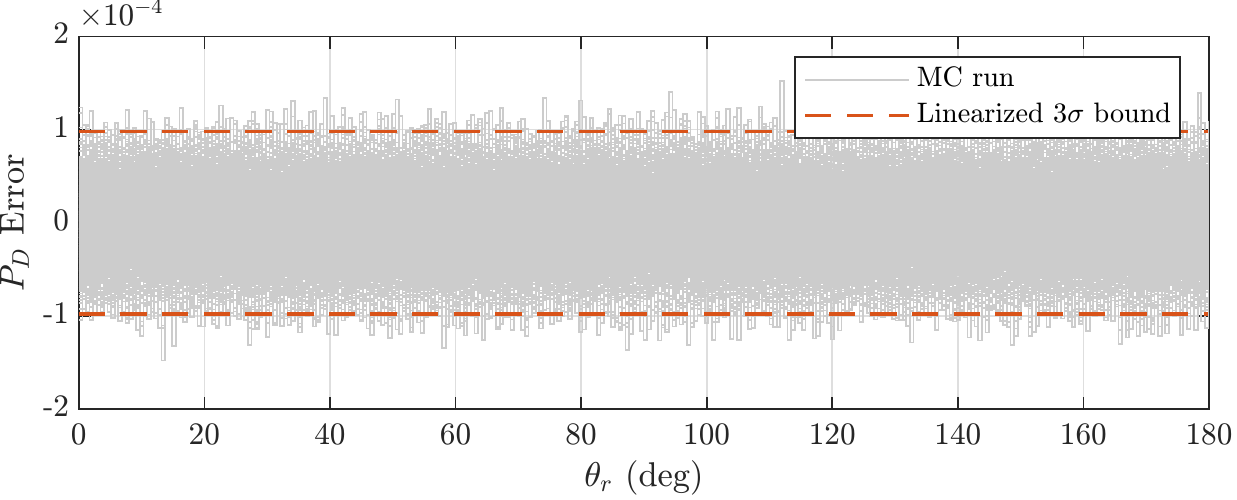}
    \caption{Monte Carlo analysis results for probability of detection with medium level of aircraft state uncertainty over the range $\theta_r = [0, 
    \; 180]$  degrees with the constant RCS model. The plots show ``hair'' lines $P_D$ (top) and $P_D$ error (bottom) for each Monte Carlo run and the 3-$\sigma_{pd}$ values from the linearized variance. \label{fig:mc_constant}}
\end{figure}

\begin{figure}
    \centering{}
    \includegraphics[width=1\columnwidth]{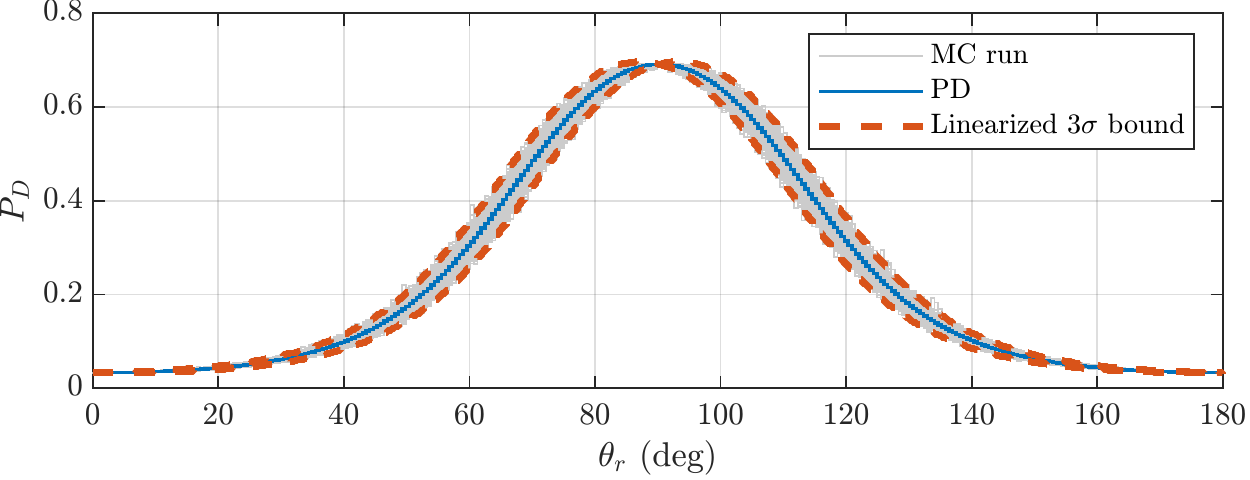}
    \includegraphics[width=1\columnwidth]{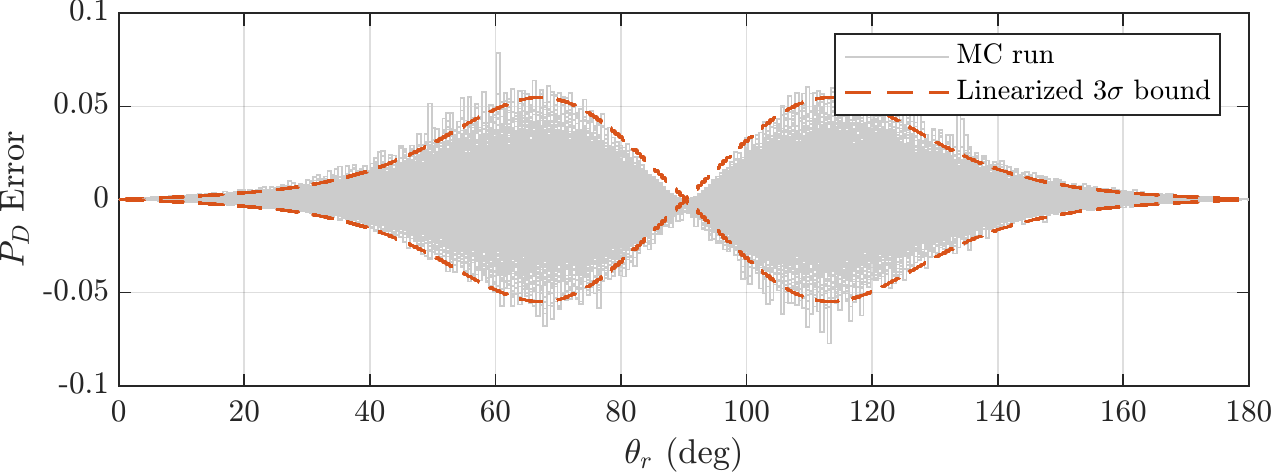}
    \caption{Monte Carlo analysis results for probability of detection with ``medium'' level of aircraft state uncertainty over the range $\theta_r = [0, \; 180]$  degrees with the ellipsoid RCS model. The plots show ``hair'' lines $P_D$ (top) and $P_D$ error (bottom) for each Monte Carlo run and the 3-$\sigma_{pd}$ values from the linearized variance. \label{fig:mc_ellipsoid}}
\end{figure}

\begin{figure}
    \centering{}
    \includegraphics[width=1\columnwidth]{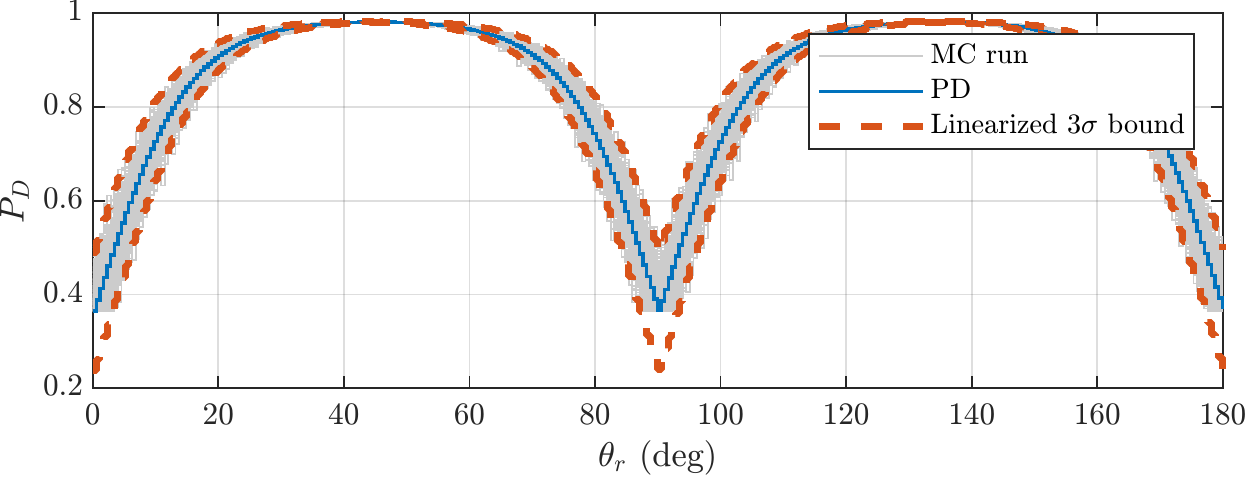}
    \includegraphics[width=1\columnwidth]{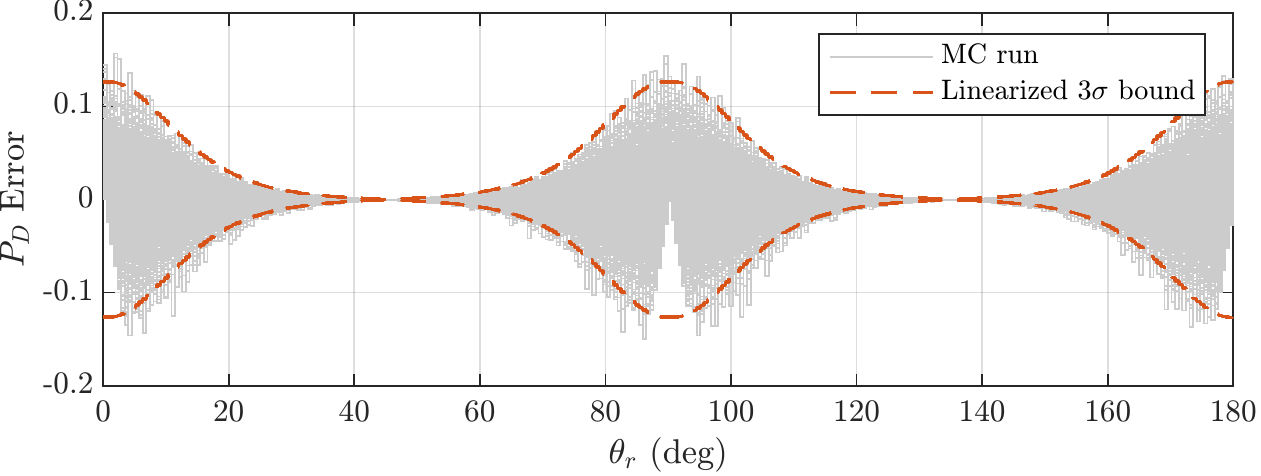}
    \caption{Monte Carlo analysis results for probability of detection with ``medium'' level of aircraft state uncertainty over the range $\theta_r = [0, \; 180]$  degrees with the spikeball RCS model. The plots show ``hair'' lines $P_D$ (top) and $P_D$ error (bottom) for each Monte Carlo run and the 3-$\sigma_{pd}$ values from the linearized variance. \label{fig:mc_spikeball}}
\end{figure}

The Monte Carlo results are provided in Figs. \ref{fig:mc_constant}-\ref{fig:mc_spikeball} which show gray lines representing $P_D$ calculated for each Monte Carlo run at a given radar azimuth angle, $\theta_r$.
\revisionD{
The standard deviation of the linearized system, $\sigma_{pd}$, is calculated using \eqref{eq:pd_sig} with $\boldsymbol{\bar{x}_a}$ and the associated 3-$\sigma_{pd}$ values are illustrated with dashed red lines.
For valid linearization, the 3-$\sigma_{pd}$ values calculated from the linearized system will be consistent with the Monte Carlo results.
This is shown graphically by the red dashed lines mostly encapsulating the the gray lines representing $P_D$ calculated for each Monte Carlo run.
} 
The following paragraphs describe the validation results for the three RCS models introduced in Section \ref{sec:rcs_models}.

The Monte Carlo result for the constant RCS model is shown in Fig. \ref{fig:mc_constant}.
\revisionD{
The top plot shows the $P_D$ result for each of the Monte Carlo runs with the $\bar{P}_D\pm$3-$\sigma_{pd}$ values from the linearized variance. The bottom figure illustrates the difference between $\bar{P}_D$ and the $P_D$ for each of the Monte Carlo runs.
}
Two observations are noteworthy.  First, the variation in $P_D$ due to uncertainty in aircraft pose is small, less than 0.1\%.  This result is expected, as the RCS is independent of vehicle attitude and the variation in vehicle position is small compared to the range, $R$. In many applications, this level of variability is insignificant, and therefore can be neglected during mission planning.  Regardless, the 3-$\sigma_{pd}$ values predicted by \eqref{eq:pd_sig} are consistent with the ensemble statistics for all radar azimuth angles.

Fig. \ref{fig:mc_ellipsoid} shows the Monte Carlo result for the ellipsoid RCS.
In contrast with the constant RCS model, $\bar{P}_D$ for the ellipsoid RCS ranges from near 0 to 0.7 as a function of the radar azimuth angle $\theta_r$. Furthermore, the bottom plot of Fig. \ref{fig:mc_ellipsoid} illustrates up to 5\% variability in $P_D$, and is a strong function of $\theta_r$.  As was the case for the constant RCS, the 3-$\sigma_{pd}$ values predicted by \eqref{eq:pd_sig} are consistent with the ensemble statistics for all azimuth angles.

Fig. \ref{fig:mc_spikeball} shows similar results for the spikeball RCS.  As in the previous case, $\bar{P}_D$ exhibits a strong dependence on the azimuth angle, ranging from 0.25 to 0.9.  For azimuth angles of 0, 90, and 180 degrees, the variation in PD is large, slightly exceeding $\pm10$\%.  As in the previous two cases, 3-$\sigma_{pd}$ values obtained via linearization appropriately bound Monte Carlo ensembles for all radar azimuth angles.  At radar azimuth angles of 0, 90, and 180 degrees, however, the Monte Carlo ensembles become biased.  This is due to the sharp corners observed at the same angles for the spikeball RCS. Regardless, the Monte Carlo ensembles are bounded by the 3-$\sigma_{pd}$ values calculated from the linearized models, resulting in a conservative estimate of $P_D$.

\begin{figure}
    \centering{}
    \includegraphics[width=1\columnwidth]{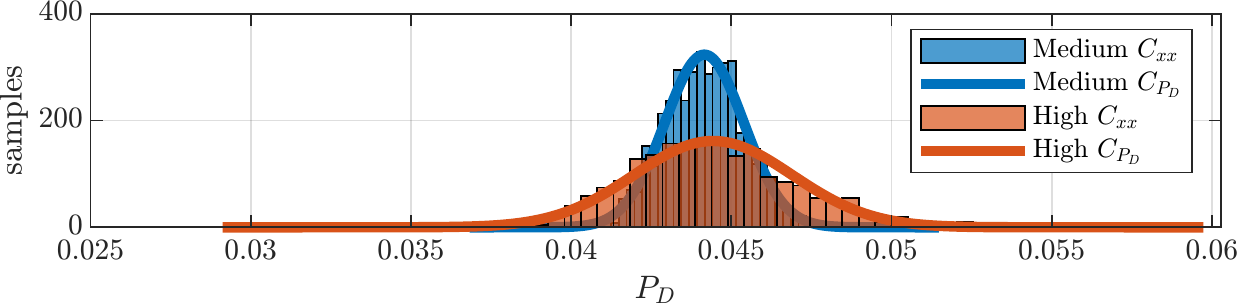}
    \includegraphics[width=1\columnwidth]{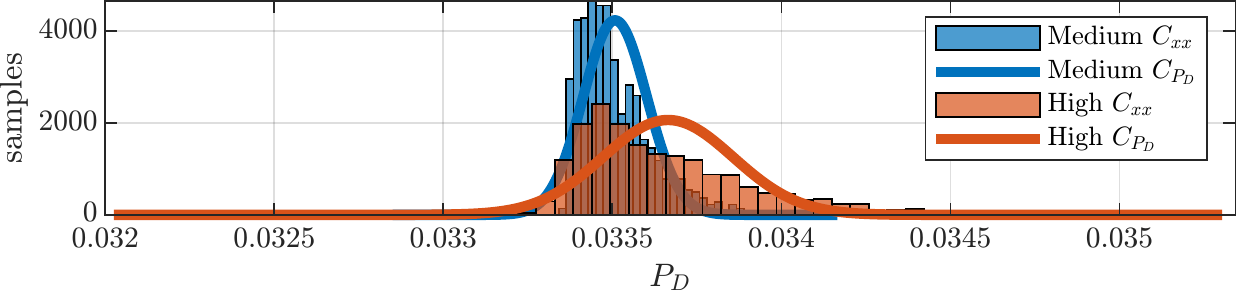}
    \caption{Linearization validation for the ellipsoid RCS with medium and high levels of state uncertainty for two values of $\theta_r$. Shows consistent linearization at $\theta_r = 20$ degrees (top), as expected, and slight linearization error at $\theta_r = 2$ degrees (bottom). \label{fig:lin_val_ellipsoid}}
\end{figure}

\begin{figure}
    \centering{}
    \includegraphics[width=1\columnwidth]{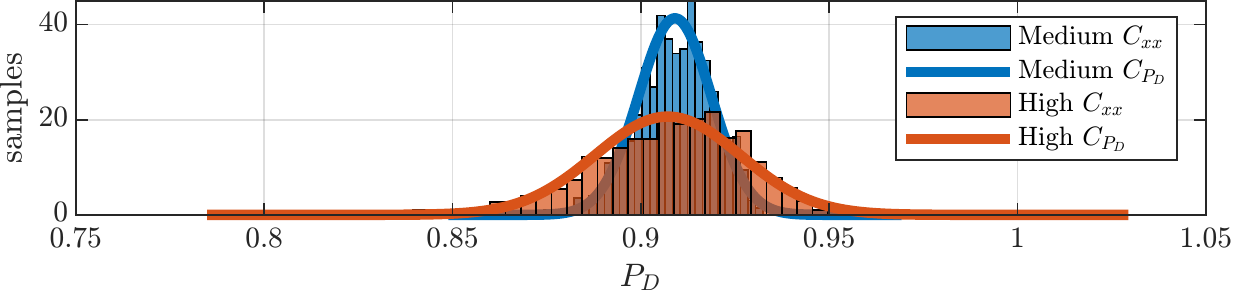}
    \includegraphics[width=1\columnwidth]{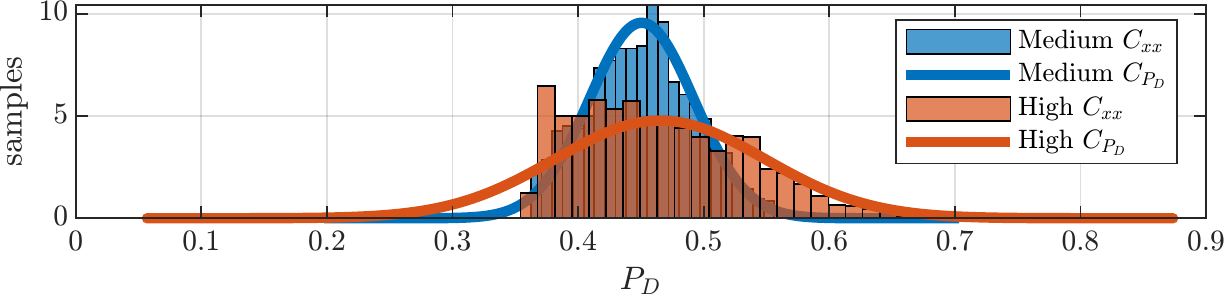}
    \caption{Linearization validation for the spikeball RCS with medium and high levels of state uncertainty for two values of $\theta_r$. Shows consistent linearization at $\theta_r = 20$ degrees (top), as expected, and slight linearization error at $\theta_r = 2$ degrees (bottom). \label{fig:lin_val_spikeball}}
\end{figure}

The previous results illustrate consistency between the variations in $P_D$ and the 3-$\sigma_{pd}$ bounds predicted by \eqref{eq:pd_sig}. The results shown in Figures \ref{fig:lin_val_ellipsoid} and \ref{fig:lin_val_spikeball} investigate the distribution of the Monte Carlo ensemble statistics for the ellipsoid and spikeball RCS values, respectively. This is accomplished by comparing the histogram of the Monte Carlo ensemble to a Gaussian distribution defined by $\bar{P}_D$ and the variance from \eqref{eq:pd_sig}.
Histogram comparisons are made for radar azimuth angles of 2 and 20 degrees, for the medium and high levels of state uncertainty.
\revisionD{
The RCS azimuth angle associated with $\theta_r=2$ degrees is near the strongest nonlinearity in the ellipsoid and spikeball RCS models and is expected to have the largest error due to linearization.
In contrast, the RCS azimuth angle associated with $\theta_r=20$ degrees is between the strong nonlinearities in the RCS models and is expected to have less error due to linearization.
}

Fig. \ref{fig:lin_val_ellipsoid} shows a histogram of the Monte Carlo results for the ellipsoid RCS. The top plot shows excellent agreement between the histogram and the corresponding Gaussian distribution for both the medium and high levels of state uncertainty with $\theta_r = 20$ degrees.  In the bottom plot, for the case of $\theta_r = 2$ degrees, a subtle negative skew develops in the histogram due to the interior corners of the Ellipsoid RCS at 0 and 180 degrees.

Fig. \ref{fig:lin_val_spikeball} shows a histogram of the Monte Carlo results for the spikeball RCS.  Similar to the ellipsoid RCS, excellent agreement is observed between the histogram and the corresponding Gaussian distribution at an azimuth angle of 20 degrees.  Also similar is the left distribution skew induced by the sharp interior corners of the spikeball RCS, especially for the high uncertainty case.

Several important conclusions are drawn from the results of this section.  First, the dependence of $P_D$ on the aircraft pose increases with the complexity of the RCS -- independent for the constant case and strongly dependent for the ellipsoid and spikeball cases.  Similarly, the variability of $P_D$ due to uncertainty in aircraft pose is substantial for ellipsoid and spikeball RCS models, $\pm5\%$ for the former and $\pm10\%$ for the latter. Furthermore, sharp changes in the RCS with respect to azimuth angle induce a skew in the distribution of $P_D$ about the nominal value.  The skew is especially prominent for the case of high state uncertainties. The 3-$\sigma_{pd}$ values predicted by \eqref{eq:pd_sig}, however, are consistent with the extents of the Monte Carlo ensembles, yielding conservative bounds for $P_D$.

\subsection{Sensitivity to State Uncertainty}
The results of the previous section served to validate the variation in $P_D$ predicted by \eqref{eq:pd_sig}. This section explores the sensitivity of $P_D$ to state uncertainty using the linearized models. This is illustrated by evaluating 3-$\sigma_{pd}$ values over a range of $\theta_r$ for the three RCS models and the three levels of state uncertainty.
Note that the nominal range for each RCS model is the same as in the previous section.

\begin{figure}
    \centering{}
    \includegraphics[width=1\columnwidth]{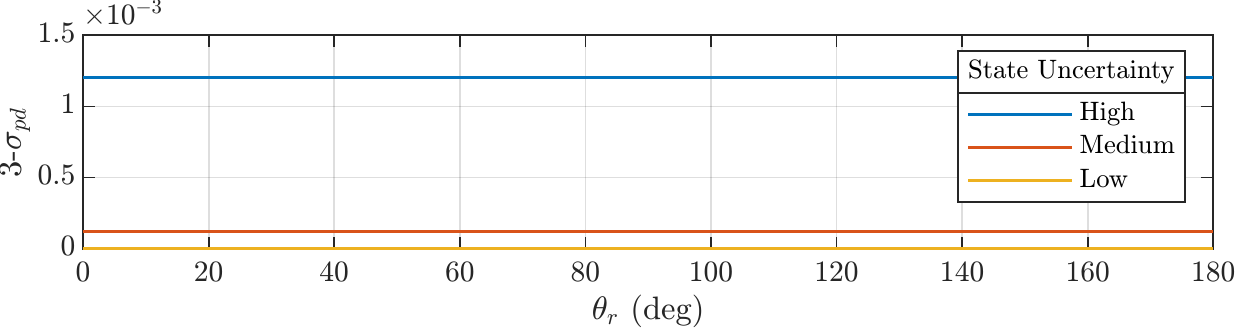}
    \caption{Probability of detection sensitivity to  state uncertainty for the constant RCS. Each line indicates the 3-$\sigma_{pd}$ values of the linearized $P_D$ statistics for the range $\theta_r = [0 \; 180]$ degrees. \label{fig:pd_3sig_const}}
\end{figure}

Fig. \ref{fig:pd_3sig_const} shows 3-$\sigma_{pd}$ of the linearized $P_D$ statistics for the constant RCS over a range of $\theta_r$ with the three levels of state uncertainty. These results indicate that the variation of $P_D$, though small, increases for higher levels of state uncertainty. Similar to the Monte Carlo results for the constant RCS model, the small variations are expected and are insignificant for many applications. 

\begin{figure}
    \centering{}
    \includegraphics[width=1\columnwidth]{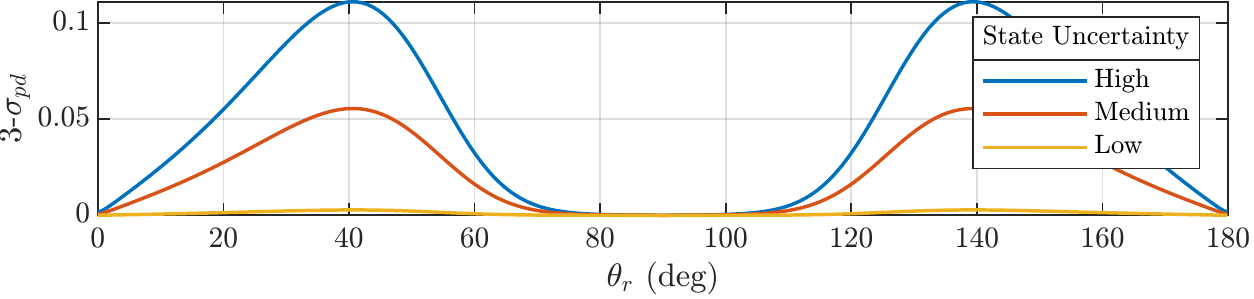}
    \caption{Probability of detection sensitivity to state uncertainty for the ellipsoid RCS. Each line indicates the 3-$\sigma_{pd}$ values of the linearized $P_D$ statistics.\label{fig:pd_3sig_ellipsoid}}
\end{figure}

\begin{figure}
    \centering{}
    \includegraphics[width=1\columnwidth]{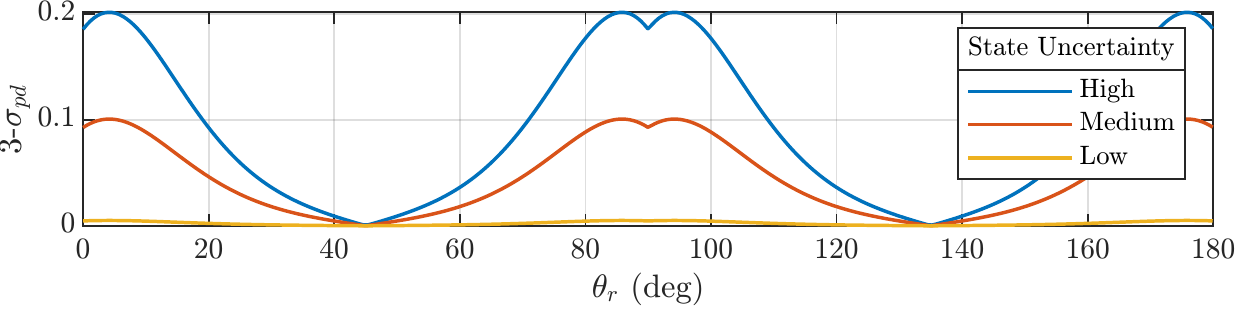}
    \caption{Probability of detection sensitivity to state uncertainty for the spikeball RCS. Each line indicates the 3-$\sigma_{pd}$ values of the linearized $P_D$ statistics.\label{fig:pd_3sig_spikeball}}
\end{figure}

Fig. \ref{fig:pd_3sig_ellipsoid} shows 3-$\sigma_{pd}$ for the ellipsoid RCS model across a range of $\theta_r$ for the three state uncertainty levels. The plot shows that 3-$\sigma_{pd}$ exceeds $0.1$ for the high level of state uncertainty. The magnitude of the sensitivity is highly dependent on the radar azimuth angle $\theta_r$. Fig. \ref{fig:pd_3sig_spikeball} shows similar results for the spikeball RCS but 3-$\sigma_{pd}$ exceeds $0.2$ for the high level of state uncertainty. 


The results in this section show that 3-$\sigma_{pd}$ varies based on the RCS model, radar azimuth angle $\theta_r$, and the state uncertainty. The  3-$\sigma_{pd}$ values are small for the constant RCS for each level of state uncertainty. However, the 3-$\sigma_{pd}$ values for the ellipsoid and spikeball RCS models are significant, especially for the medium and high levels of state uncertainty. This indicates that considering the state uncertainty for the ellipsoid and spikeball RCS models provides significant value, especially for scenarios with large state uncertainty. Failing to consider the state uncertainty in these scenarios will result in detection probabilities that are higher than expected.

\section{Conclusion \label{sec:Conclusion}}
In planning the path of an aircraft, operators must ensure that the probability of being detected by radar systems stays below mission-specified levels. The probability of detection $P_D$ is commonly considered a deterministic value that is a function of the aircraft pose, radar position, and radar parameters. This approach fails to consider the variability in $P_D$ induced by uncertainty in such parameters. A source of uncertainty in mission planning is in the aircraft pose which is often modeled as a Gaussian random vector with a mean and covariance. This paper presents a method for estimating the variability of $P_D$ due to uncertainty in the aircraft pose.

The method presented in this work provides a first-order approximation of the variations in $P_D$. This is accomplished by linearizing an expression for $P_D$ with respect to the aircraft pose.  The resulting linear model is used to determine the 3-$\sigma$ variability of $P_D$ due to uncertainty in the aircraft pose. As part of the linearization, the necessary partial derivatives are derived for three radar cross section models--constant, ellipsoid, and spikeball.

The linearization of $P_D$ with respect to the aircraft state is validated using a Monte Carlo analysis for three levels of aircraft pose uncertainty and the three radar cross section models. Despite small biases present in the Monte Carlo ensemble due to sharp changes in the RCS model, the 3-$\sigma$ values predicted by the linear model appropriately bounded the variations in $P_D$ due to uncertainty in the aircraft pose.

The sensitivity of $P_D$ to aircraft pose uncertainty was also explored in this paper. The results showed that the magnitude of the 3-$\sigma$ bound is significant for the ellipsoid and simple spikeball RCS models -- over $0.1$ and $0.2$, respectively for a high level of state uncertainty. These variations are significant and must be considered in mission planning efforts. Failing to consider the aircraft pose uncertainty will result in detection probabilities that are higher than expected.

\bibliographystyle{IEEEtran}
\bibliography{full_bib}

\begin{thebibliography}{10}
\providecommand{\url}[1]{#1}
\csname url@samestyle\endcsname
\providecommand{\newblock}{\relax}
\providecommand{\bibinfo}[2]{#2}
\providecommand{\BIBentrySTDinterwordspacing}{\spaceskip=0pt\relax}
\providecommand{\BIBentryALTinterwordstretchfactor}{4}
\providecommand{\BIBentryALTinterwordspacing}{\spaceskip=\fontdimen2\font plus
\BIBentryALTinterwordstretchfactor\fontdimen3\font minus
  \fontdimen4\font\relax}
\providecommand{\BIBforeignlanguage}[2]{{%
\expandafter\ifx\csname l@#1\endcsname\relax
\typeout{** WARNING: IEEEtran.bst: No hyphenation pattern has been}%
\typeout{** loaded for the language `#1'. Using the pattern for}%
\typeout{** the default language instead.}%
\else
\language=\csname l@#1\endcsname
\fi
#2}}
\providecommand{\BIBdecl}{\relax}
\BIBdecl

\bibitem{Ceccarelli_micro_uav}
N.~Ceccarelli, J.~J. Enright, E.~Frazzoli, S.~J. Rasmussen, and C.~J.
  Schumacher, ``Micro uav path planning for reconnaissance in wind,'' in
  \emph{2007 American Control Conference}, 2007, pp. 5310--5315.

\bibitem{larson_path_nodate}
R.~Larson, M.~Pachter, and M.~Mears, ``Path {Planning} by {Unmanned} {Air}
  {Vehicles} for {Engaging} an {Integrated} {Radar} {Network},'' in
  \emph{{AIAA} {Guidance}, {Navigation}, and {Control} {Conference} and
  {Exhibit}}.\hskip 1em plus 0.5em minus 0.4em\relax American Institute of
  Aeronautics and Astronautics, 2001.

\bibitem{xiao-wei_path_2010}
F.~Xiao-wei, L.~Zhong, and G.~Xiao-guang, ``Path {Planning} for {UAV} in
  {Radar} {Network} {Area},'' in \emph{2010 {Second} {WRI} {Global} {Congress}
  on {Intelligent} {Systems}}, vol.~3, Dec. 2010, pp. 260--263.

\bibitem{kabamba_optimal_2012}
P.~T. Kabamba, S.~M. Meerkov, and F.~H.~Z. III,
  ``\BIBforeignlanguage{en}{Optimal {Path} {Planning} for {Unmanned} {Combat}
  {Aerial} {Vehicles} to {Defeat} {Radar} {Tracking}},''
  \emph{\BIBforeignlanguage{en}{Journal of Guidance, Control, and Dynamics}},
  May 2012.

\bibitem{mcfarland_motion_1999}
M.~McFarland, R.~Zachery, and B.~Taylor, ``Motion planning for reduced
  observability of autonomous aerial vehicles,'' in \emph{Proceedings of the
  1999 {IEEE} {International} {Conference} on {Control} {Applications} ({Cat}.
  {No}.{99CH36328})}, vol.~1, Aug. 1999, pp. 231--235 vol. 1.

\bibitem{bortoff_path_2000}
S.~Bortoff, ``Path planning for {UAVs},'' in \emph{Proceedings of the 2000
  {American} {Control} {Conference}. {ACC} ({IEEE} {Cat}. {No}.{00CH36334})},
  vol.~1, Jun. 2000, pp. 364--368 vol.1, iSSN: 0743-1619.

\bibitem{chandler_uav_2000}
P.~Chandler, S.~Rasmussen, and M.~Pachter, ``\BIBforeignlanguage{en}{{UAV}
  cooperative path planning},'' in \emph{\BIBforeignlanguage{en}{{AIAA}
  {Guidance}, {Navigation}, and {Control} {Conference} and {Exhibit}}}.\hskip
  1em plus 0.5em minus 0.4em\relax Dever,CO,U.S.A.: American Institute of
  Aeronautics and Astronautics, Aug. 2000.

\bibitem{pachter_optimal_2001}
M.~Pachter and J.~Hebert, ``Optimal aircraft trajectories for radar exposure
  minimization,'' in \emph{Proceedings of the 2001 {American} {Control}
  {Conference}. ({Cat}. {No}.{01CH37148})}, vol.~3, Jun. 2001, pp. 2365--2369
  vol.3, iSSN: 0743-1619.

\bibitem{moore_radar_2002}
F.~Moore, ``Radar cross-section reduction via route planning and intelligent
  control,'' \emph{IEEE Transactions on Control Systems Technology}, vol.~10,
  no.~5, pp. 696--700, Sep. 2002.

\bibitem{jun_path_2003}
M.~Jun and R.~D’Andrea, ``\BIBforeignlanguage{en}{Path {Planning} for
  {Unmanned} {Aerial} {Vehicles} in {Uncertain} and {Adversarial}
  {Environments}},'' in \emph{\BIBforeignlanguage{en}{Cooperative {Control}:
  {Models}, {Applications} and {Algorithms}}}, ser. Cooperative {Systems},
  S.~Butenko, R.~Murphey, and P.~M. Pardalos, Eds.\hskip 1em plus 0.5em minus
  0.4em\relax Boston, MA: Springer US, 2003, pp. 95--110.

\bibitem{savage_strapdown_2000}
P.~G. Savage, \emph{\BIBforeignlanguage{English}{Strapdown analytics}}.\hskip
  1em plus 0.5em minus 0.4em\relax Maple Plain, Minn.: Strapdown Associates,
  2000.

\bibitem{farrell_aided_2008}
J.~Farrell, \emph{Aided {Navigation}: {GPS} with {High} {Rate} {Sensors}},
  1st~ed.\hskip 1em plus 0.5em minus 0.4em\relax USA: McGraw-Hill, Inc., 2008.

\bibitem{grewal_global_2020}
M.~S. Grewal, A.~P. Andrews, and C.~G. Bartone,
  \emph{\BIBforeignlanguage{en}{Global {Navigation} {Satellite} {Systems},
  {Inertial} {Navigation}, and {Integration}}}.\hskip 1em plus 0.5em minus
  0.4em\relax John Wiley \& Sons, Jan. 2020.

\bibitem{marcum_statistical_1960}
J.~Marcum, ``A statistical theory of target detection by pulsed radar,''
  \emph{IRE Transactions on Information Theory}, vol.~6, no.~2, pp. 59--267,
  Apr. 1960.

\bibitem{swerling_probability_1954}
P.~Swerling, ``\BIBforeignlanguage{en}{Probability of {Detection} for
  {Fluctuating} {Targets}},'' RAND Corporation, Tech. Rep., Jan. 1954.

\bibitem{mahafza_matlab_2003}
B.~R. Mahafza and A.~Elsherbeni, \emph{\BIBforeignlanguage{en}{{MATLAB}
  {Simulations} for {Radar} {Systems} {Design}}}.\hskip 1em plus 0.5em minus
  0.4em\relax CRC Press, Dec. 2003.

\bibitem{north_analysis_1963}
D.~North, ``An {Analysis} of the factors which determine signal/noise
  discrimination in pulsed-carrier systems,'' \emph{Proceedings of the IEEE},
  vol.~51, no.~7, pp. 1016--1027, Jul. 1963.

\bibitem{beard_randy_small_2012}
{Beard, Randy} and {McLain, Timothy}, \emph{Small {Unmanned} {Aircraft}
  {Theory} and {Practice}}, 2012.

\bibitem{weibel_safety_2005}
\BIBentryALTinterwordspacing
R.~E. Weibel, ``\BIBforeignlanguage{eng}{Safety considerations for operation of
  different classes of unmanned aerial vehicles in the {National} {Airspace}
  {System}},'' Thesis, Massachusetts Institute of Technology, 2005. [Online].
  Available: \url{https://dspace.mit.edu/handle/1721.1/30364}
\BIBentrySTDinterwordspacing

\end{thebibliography}

\appendices{}

\section{Radar Cross Section Linearization \label{ap:rcs_linearization}}
The following paragraphs will define the partial derivatives of the three radar detection models presented in Section \ref{sec:rcs_models}. The ellipsoid and simple spikeball RCS models are functions of the RCS azimuth and elevation angles from the radar detection vector. The partial derivatives of these expressions are derived as follows
\begin{eqnarray}
    \partiald{\lambda}{\boldsymbol{x_a}} &=& \partiald{\lambda}{\boldsymbol{p_r^b}} \partiald{\boldsymbol{p_r^b}}{\boldsymbol{x_a}} \\
    \partiald{\phi}{\boldsymbol{x_a}} &=& \partiald{\phi}{\boldsymbol{p_r^b}} \partiald{\boldsymbol{p_r^b}}{\boldsymbol{x_a}}
\end{eqnarray}
where 
\begin{equation}
    \partiald{\boldsymbol{p_r^b}}{\boldsymbol{x_a}} = \begin{bmatrix} \partiald{\boldsymbol{p_r^b}}{\boldsymbol{p_a^n}} & \partiald{\boldsymbol{p_r^b}}{\boldsymbol{\Theta_a}} \end{bmatrix}.
\end{equation}
The partial derivative of the azimuth and elevation angle with respect to $\boldsymbol{p_r^b}$ are given by
\begin{eqnarray}
    \partiald{\lambda}{\boldsymbol{p_r^b}} &=& \begin{bmatrix} \frac{-p_{ry}}{p_{rx}^2 + p_{ry}^2} & 
    \frac{p_{rx}}{p_{rx}^2 + p_{ry}^2} &
    0 \end{bmatrix} \\
    \partiald{\phi}{\boldsymbol{p_r^b}} &=& \begin{bmatrix} \frac{- p_{rx} p_{rz}}{\alpha} & 
    \frac{- p_{ry} p_{rz}}{\alpha}  &
    \frac{\sqrt{p_{rx}^2 + p_{ry}^2}}{p_{rx}^2 + p_{ry}^2 + p_{rz}^2} \end{bmatrix}
\end{eqnarray}
where
\begin{equation}
    \alpha = \left(p_{rx}^2 + p_{ry}^2 + p_{rz}^2\right)\sqrt{p_{rx}^2 + p_{ry}^2}.
\end{equation}
The partial derivative of $\boldsymbol{p_r^b}$ with respect to the aircraft position in the NED frame is given by
\begin{equation}
    \partiald{\boldsymbol{p_r^b}}{\boldsymbol{p_a^n}} = -R_n^b
\end{equation}
where $R_n^b$ is defined in \eqref{eq:DCM_zyx}. The partial derivative of $\boldsymbol{p_r^b}$ with respect to the aircraft orientation is given by
\begin{equation}
    \partiald{\boldsymbol{p_r^b}}{\boldsymbol{\Theta_a}} = \begin{bmatrix} \partiald{\boldsymbol{p_r^b}}{\boldsymbol{\Theta_a}}_{11} & \partiald{\boldsymbol{p_r^b}}{\boldsymbol{\Theta_a}}_{12} & \partiald{\boldsymbol{p_r^b}}{\boldsymbol{\Theta_a}}_{13} \\
        \partiald{\boldsymbol{p_r^b}}{\boldsymbol{\Theta_a}}_{21} & \partiald{\boldsymbol{p_r^b}}{\boldsymbol{\Theta_a}}_{22} & \partiald{\boldsymbol{p_r^b}}{\boldsymbol{\Theta_a}}_{23} \\
        \partiald{\boldsymbol{p_r^b}}{\boldsymbol{\Theta_a}}_{31} & \partiald{\boldsymbol{p_r^b}}{\boldsymbol{\Theta_a}}_{32} & \partiald{\boldsymbol{p_r^b}}{\boldsymbol{\Theta_a}}_{33} \end{bmatrix}
\end{equation}
where
\begin{eqnarray}
    \partiald{\boldsymbol{p_r^b}}{\boldsymbol{\Theta_a}}_{11} &=& -(C{\phi_a} S{\psi_a} - C{\psi_a} S{\phi_a} S{\theta_a})p_{\Delta d} \nonumber \\
            && \quad -(S{\phi_a} S{\psi_a} + C{\phi_a} C{\psi_a} S{\theta_a})p_{\Delta e} \\
    \partiald{\boldsymbol{p_r^b}}{\boldsymbol{\Theta_a}}_{12} &=& C{\psi_a} S{\theta_a} p_{\Delta n} - C{\psi_a} C{\theta_a} S{\phi_a} p_{\Delta e} \nonumber  \\
            && \quad - C{\phi_a} C{\psi_a} C{\theta_a} p_{\Delta d} \\
    \partiald{\boldsymbol{p_r^b}}{\boldsymbol{\Theta_a}}_{13} &=& (C{\phi_a} C{\psi_a} + S{\phi_a}S{\psi_a}S{\theta_a})p_{\Delta e} \nonumber  \\
            && \quad - (C{\psi_a} S{\phi_a} - C{\phi_a} S{\psi_a} S{\theta_a}) p_{\Delta d} \nonumber  \\
            && \quad + C{\theta_a} S{\psi_a} p_{\Delta n} \\
    \partiald{\boldsymbol{p_r^b}}{\boldsymbol{\Theta_a}}_{21} &=& (C{\phi_a} C{psi_a} + S{\phi_a} S{\psi_a} S{\theta_a}) p_{\Delta d} \nonumber \\
        && \quad + (C{\psi_a} S{\phi_a} - C{\phi_a} S{\psi_a} S{\theta_a}) p_{\Delta e} \\
    \partiald{\boldsymbol{p_r^b}}{\boldsymbol{\Theta_a}}_{22} &=& S{\psi_a} S{\theta_a} p_{\Delta n} - C{\phi_a} C{\theta_a} S{\psi_a} p_{\Delta d} \nonumber \\
        && \quad - C{\theta_a} S{\phi_a} S{\psi_a} p_{\Delta e} \\
    \partiald{\boldsymbol{p_r^b}}{\boldsymbol{\Theta_a}}_{23} &=& (C{\phi_a} S{\psi_a} - C{\psi_a}S{\phi_a}S{\theta_a})p_{\Delta e} \nonumber \\
        && \quad - (S{\phi_a} S{\psi_a} + C{\phi_a}C{\psi_a}S{\theta_a}) p_{\Delta d} \nonumber \\
        && \quad - C{\psi_a} C{\theta_a} p_{\Delta n} \\
    \partiald{\boldsymbol{p_r^b}}{\boldsymbol{\Theta_a}}_{31} &=& C{\theta_a} S{\phi_a} p_{\Delta d} - C{\phi_a} C{\theta_a} p_{\Delta e} \\
    \partiald{\boldsymbol{p_r^b}}{\boldsymbol{\Theta_a}}_{32} &=& C{\theta_a} p_{\Delta n} + C{\phi_a} S{\theta_a} p_{\Delta d} + S{\phi_a} S{\theta_a} p_{\Delta e}   \\
    \partiald{\boldsymbol{p_r^b}}{\boldsymbol{\Theta_a}}_{32} &=& 0
\end{eqnarray}
and
\begin{eqnarray}
    \boldsymbol{p_\Delta^n} &=& \boldsymbol{p_r^n}-\boldsymbol{p_a^n} \\
        &=& \begin{bmatrix} p_{\Delta n} & p_{\Delta e} & p_{\Delta d} \end{bmatrix}^\intercal.
\end{eqnarray}
The partial derivatives of the RCS models with respect to $\lambda$ and $\phi$ are defined in the following subsections.

\subsection{Constant RCS}
The first RCS model is the constant radar cross section. This model is independent of the aircraft state so
\begin{equation}
\frac{\partial\sigma_{rc}}{\partial\boldsymbol{x_{a}}} = \boldsymbol{0_{1\times6}}.
\end{equation} 

\subsection{Ellipsoid RCS}
The second RCS model is the ellipsoid. The partial derivative of the ellipsoid RCS model with respect to the aircraft state is given by
\begin{equation}
    \partiald{\sigma_{re}}{\boldsymbol{x_a}} = \partiald{\sigma_{re}}{\lambda} \partiald{\lambda}{\boldsymbol{x_a}} + \partiald{\sigma_{re}}{\phi} \partiald{\phi}{\boldsymbol{x_a}}.
\end{equation}
The partial derivatives of the ellipsoid RCS model with respect to the RCS azimuth and elevation angles are given by 
\begin{eqnarray}
    \partiald{\sigma_{re}}{\lambda} &=& \frac{-2 \pi (abc)^2\sin(2\lambda)\kappa}{D^3} \\
    \partiald{\sigma_{re}}{\phi} &=& \frac{-2 \pi (abc)^2\left(b^2-a^2\right)\sin(\lambda)^2\sin(2\phi)}{D^3}
\end{eqnarray}
where
\begin{equation}
    \kappa = \left(a^2\cos(\phi)^2+b^2\sin(\phi)^2+c^2\right)
\end{equation}
and
\begin{equation}
    \textrm{D} = \left(a \,\sin\lambda \, \sin \phi\right)^{2}+\left(b \,\sin\lambda\, \sin\phi\right)^{2}+\left(c\, \cos\phi\right)^{2}.
\end{equation}

\subsection{Simple Spikeball RCS}
The third radar cross section model is the simple spikeball. The partial derivative of the simple spikeball RCS model in \eqref{eq:spikeball_rcs}
with respect to the aircraft state is
\begin{equation}
    \partiald{\sigma_{rs}}{\boldsymbol{x_a}} = \partiald{\sigma_{rs}}{\lambda} \partiald{\lambda}{\boldsymbol{x_a}}
\end{equation}
where 
\begin{equation}
    \partiald{\sigma_{rs}}{\lambda} = \frac{n}{2}a_s\cos\left(\frac{n}{2}\lambda\right)\text{{sign}}\left(a_s\sin\left(\frac{n}{2}\lambda\right)\right).
\end{equation}

\end{document}